\pdfoutput=1

\documentclass[11pt]{article}

\usepackage{geometry}
\usepackage{graphicx}
\usepackage{amssymb}
\usepackage{epstopdf}
\usepackage{verbatim,setspace}
\usepackage[numbers,sort&compress]{natbib}
\usepackage{color}\definecolor{DkRed}{cmyk}{0,.5,.5,.3}

\def\NLl{L_{\rm loop}}

\def\sref#1{Sect.~\ref{s:#1}}

\def\bpunit{\ensuremath{\mathrm{bp}}}
\def\nmunit{\ensuremath{\mathrm{nm}}}

\title{Concentration and Length Dependence of DNA Looping in Transcriptional Regulation}

\begin{document}

\author{Lin Han$^{1}$, Hernan G.\   Garcia$^{2}$, Seth Blumberg$^{1}$, Kevin B. Towles$^{3}$,\\  John F.\
Beausang$^{3}$,
Philip C.\  Nelson$^{3}$,  Rob Phillips$^{1*}$\\
$^{1}$Department of Applied Physics,
California Institute of\\
Technology, Pasadena CA 91125;\\
$^{2}$Department of  Physics,
California Institute of\\
Technology, Pasadena CA 91125;\\
$^{3}$Department of Physics and Astronomy, University of  Pennsylvania, \\ Philadelphia PA 19104\\
$^{*}$Corresponding author: {\tt phillips@pboc.caltech.edu}}
\maketitle

\begin{abstract}
{\bf BACKGROUND.} In many cases, transcriptional regulation involves
the binding of transcription factors at sites on the DNA that are
not immediately adjacent to the promoter of interest. This action at
a distance is often mediated by the formation of DNA loops: Binding
at two or more sites on the DNA results in the formation of a loop,
which can bring the transcription factor into the immediate
neighborhood of the relevant promoter.   These processes are
important in settings ranging from the historic bacterial examples
(bacterial metabolism and the lytic-lysogeny decision in
bacteriophage), to the modern concept of gene regulation to
regulatory processes central to pattern formation during development
of multicellular organisms.

{\bf METHODOLOGY/PRINCIPAL FINDINGS.} Though there have been a
variety of insights into the combinatorial aspects of
transcriptional control, the mechanism of DNA looping as an agent of
combinatorial control in both prokaryotes and eukaryotes remains
unclear.    We use single-molecule techniques to dissect DNA looping
in the {\it lac} operon. In particular, we measure the propensity
for DNA looping by the Lac repressor as a function of the
concentration of repressor protein and as a function of the distance
between repressor binding sites.

{\bf CONCLUSIONS/SIGNIFICANCE.} As with earlier single-molecule
studies, we find (at least) two distinct looped states and
demonstrate that the presence of these two states depends  both upon
the concentration of repressor protein and the distance between the
two repressor binding sites.   We find that loops form even at
interoperator spacings considerably shorter than the DNA persistence
length, without the intervention of any other proteins to prebend
the DNA.   The concentration measurements also permit us to use a
simple statistical mechanical model of DNA loop formation  to
determine the free energy of DNA looping, or equivalently, the
$J$-factor for looping.
\end{abstract}
\newpage
\renewcommand{\baselinestretch}{1.1}\normalsize

\section{Introduction}
The  biological significance of DNA is primarily attributed to the
information implicit in its  sequence. Still,  there are a wide
range of processes for which DNA's physical basis as a stiff polymer
also matters~\citep{Garcia2007a}.  For example, the packaging of DNA
into nucleosomes appears to select for sequence motifs that are
particularly flexible~\citep{Segal2006,Cloutier2004}.   In the
setting of transcriptional regulation, there are a  host of
regulatory architectures both in prokaryotes and eukaryotes which
require the interaction of sequences on the DNA that are not
adjacent~\citep{Adhya1989, Schleif1992, Matthews1992, Zeller1995}.
These interactions are mediated by DNA-binding proteins, which have
to deform the DNA.  Of late, it has become possible to perform
genome-wide surveys to determine the entirety of looped
configurations induced by a given protein~\citep{Loh2006,Wei2006}.
In eukaryotes, action of transcription factors over long distances
seems the rule rather than the exception.  One of the most
transparent examples of DNA looping is in bacteria where some
repressors and activators can bind at two sites simultaneously,
resulting in a DNA loop.  This effect was first elucidated in the
context of the arabinose operon~\citep{Dunn1984}.  It is an amusing
twist of history that the two regulatory motifs considered by Jacob
and Monod, namely, the  switch  that makes the decision between the
lytic and lysogenic pathways after phage
infection~\citep{Ptashne2004} and the decision making apparatus
associated with lactose digestion in
bacteria~\citep{Schleif1992,Muller-Hill1996}, both involve DNA
looping as well.

To understand the physical mechanism of the biological action at a
distance revealed by DNA looping, it is necessary to bring both {\it
in vitro} and {\it in vivo} experiments as well as theoretical
analyses to bear on this important problem. Over the last few
decades there have been a series of impressive and beautiful
experiments from many quarters that inspired our own work. In the
{\it in vivo} context, it is especially the work of M{\"u}ller-Hill
and coworkers that demonstrates the intriguing quantitative
implications of DNA looping for regulation~\citep{Muller1996}.  In
their experiments, they tuned the length of the DNA loop in one base
pair increments and measured the resulting repression.     More
recently, these experiments have been performed with mutant
bacterial strains that were deficient in architectural proteins such
as HU, IHF and H-NS~\citep{Becker2005,Becker2007}.    On the {\it in
vitro} side, single molecule experiments using the tethered-particle
method
\citep{Schafer1991,Yin1994,Vanzi2003,Pouget2004,Blumberg2005,Pouget2006,Broek2006,TolicNorrelykke2006,Guerra2007}
have also contributed significantly~\citep{Finzi1995, Vanzi2006,
Zurla2006,Wong2007,Zurla2007,Normanno2008}. The idea of these
experiments is to tether a piece of DNA to a microscope cover slip
with a bead attached to the end.  The DNA construct has the relevant
binding sites (operators) for the protein of interest along the DNA
and when one of these proteins binds, it shortens the length of the
tether. As a result of the shorter tether, the Brownian motion of
the bead is reduced. Hence, the size of the random excursions of the
bead serves as a reporter for the status of the DNA molecule (i.e.
looped or unlooped, DNA-binding protein present or not).

In addition to single-molecule studies, {\it in vitro} biochemical assays have
also shed important light on the interactions between transcription
factors and their DNA targets.  Both filter binding assays and electrophoretic mobility shift assays have been widely used to study how variables dictating DNA mechanics such as length and degree of supercoiling,  alter the  looping process~\citep{Kramer1987,Hsieh1987, Kramer1988,Whitson1987b,Borowiec1987}.

One of the missing links in the experimental elucidation of these
problems is systematic, single-molecule experiments which probe  the
length, repressor concentration and sequence dependence of DNA
looping. Such experiments will complement earlier {\it in vivo}
work, which has already demonstrated how DNA length and repressor
concentration alter repression~\citep{Muller1996}. Our view is that
such systematic experiments will help clarify the way in which both
length and sequence contribute to the probability of DNA looping,
and begin to elucidate the mechanisms whereby transcription factors
act over long genomic distances.  Further, such experiments can
begin to shed light on broader questions of regulatory architecture
and the significance of operator placement to transcriptional
control.   To that end, we have carried out experiments that probe
the DNA looping process over a range of concentrations of repressor
protein and for a series of different loop lengths. In addition,
intrigued by the sequence preferences observed in nucleosomal DNA,
we have made looping constructs in which these highly bendable
nucleosomal sequences are taken out of their natural eukaryotic
context and are inserted between the operators that serve as binding
sites for the Lac repressor (the results of those experiments will
be reported elsewhere).  The point of this exercise is to see how
the looping probability depends upon these tunable parameters,
namely, length, repressor concentration and sequence.

Our  key results are: (1)  The concentration dependence of looping
as a function of repressor concentration (a ``titration'' curve) can
be described by a simple equilibrium statistical-mechanics model of
transcription factor-DNA interactions. The model predicts a
saturation effect, which agrees with our experimental observations.
(2) By measuring this effect, we were able to isolate the free
energy change of looping (that is, separate it from the binding free
energy change), obtaining an experimental measurement  of its value
for a range of different lengths in an uncluttered, {\it in vitro},
setting. (3) Systematic measurement of looping free energy as a
function of interoperator spacing hints at the same modulations seen
in analogous { \it in vitro} work on
cyclization~\citep{Cloutier2004,Cloutier2005}, and {\it in vivo}
work on repression~\citep{Muller1996,Becker2005}. (4) Clear
experimental signature of multiple looped states, consistent with
theory expectations~\citep{Zhang2006, Swigon2006, Geanacopoulos2001,
Balaeff2006} and other recent
experiments~\citep{Wong2007,Normanno2008}. In the remainder of the
paper, we describe a series of experiments that examine both the
length and concentration  dependence of DNA looping induced by the
Lac repressor. A companion paper gives extensive details about our
theoretical calculations \citep{Towles2008}.

\section{Results}

As argued above, one of our central concerns in performing these
experiments was to have sufficient, systematic data to make it
possible to carry out a thorough analysis of the interplay between
theories of transcriptional regulation (and DNA
looping)~\citep{Ackers1982,Buchler2003a,Bintu2005a,Bintu2005b}, and
experiment. To that end, we have carried out a series of DNA looping
experiments using the tethered-particle method~\citep{Finzi1995} for
loop lengths ranging from 300 to 310~bp in one base pair increments
as well as several representative examples for lengths below 100~bp.
The experiments described here use DNA constructs harboring  two
different operators, symmetric operator $Oid$ and primary natural
operator $O1$ as Lac repressor binding sites.   In addition, we have
explored how the looping trajectories depend upon the concentration
of Lac repressor. The particular experimental details are described
in the ``Materials and Methods'' section.

\begin{figure}
\begin{center}
\includegraphics[width=5in]{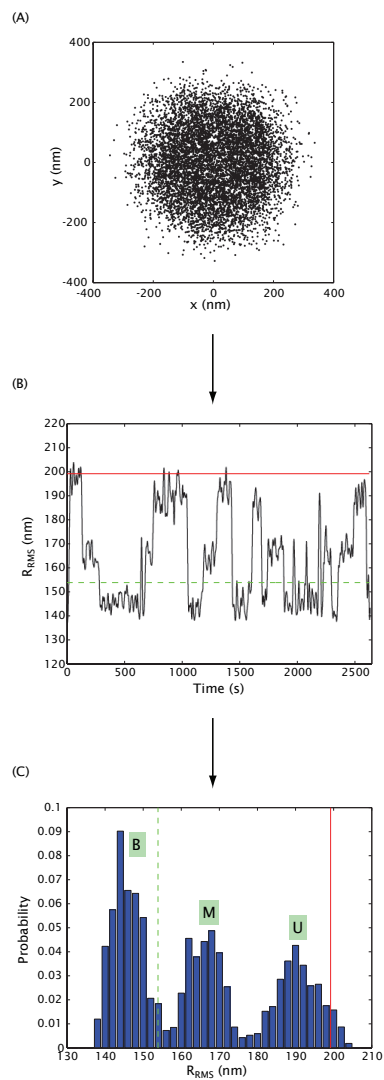}
\caption{\small{Different representations of TPM data.  (A) Scatter plot of drift-corrected positional data.
Each dot corresponds to the instantaneous projected position of the bead at a particular instant in time.
(B) Running average of Gaussian filtered RMS motion over an effective window of 4~seconds.  $R$ is the distance from
the bead center (dots in panel (A)) to the tether attachment point (centroid of all dots in panel (A)).
Red (solid)  and green (dashed)  lines represent naively expected motion, based on calibration
measurements~\citep{calibration},
for 901bp DNA and an imagined DNA for which
305+20.5~bp (the center to center distance between operators) are subtracted off of the full
length 901~bp tether. (Fig.~\ref{fig:PNCombined} gives a more precise prediction of the expected excursions in looped states.)
(C) Histogram of the RMS motion.
Different peaks correspond to looped (labeled B, bottom, and M, middle) or unlooped (labeled U) states.  The DNA used
here is pUC305L1 (see Materials and Methods section) with 100~pM Lac repressor.   A detailed discussion of how
to go from microscopy images of beads to traces and histograms like those shown here
is given in the Supplementary Materials.}
}
\label{fig:trajectory}
\end{center}
\end{figure}
A typical experimental trace resulting from these measurements  is
shown in fig.~\ref{fig:trajectory}. (Representative examples of
experimental traces from all of the lengths and concentrations
considered throughout the paper as well as examples of rejected
traces are shown in the Supplementary Material.) As seen in the
figure, as with other recent work~\citep{Wong2007,Normanno2008},
there are clearly two distinct looped states as seen both in the
trajectory and the histogram. Control experiments with one of the
two binding sites  removed show only the highest peak, which further
supports the idea that the two lower peaks indeed indicate looped
configurations. One hypothesis is that these two looped states
correspond to two different configurations of the Lac repressor
molecule and its attendant DNA, which we will refer to as the
``open'' and ``closed'' configurations.  Direct interconversion
between the  two looped species suggested the two distinct looped
states are indeed due to different conformations of Lac repressor
protein~\citep{Wong2007}.  An alternate hypothesis is that the two
peaks reflect different DNA topologies~\citep{Friedman1995,
Mehta1999, Semsey2004}. Although this hypothesis does not obviously
accommodate the apparent observation of direct interconversion,
nevertheless we will present data from Monte Carlo simulations of
DNA chain conformations that show that it \textit{can}
quantitatively explain
 the observed multi-peak structure observed in the
data.

\subsection{Concentration dependence}

In order to extract quantities such as the free energy of looping
associated with repressor binding (or equivalently, a
$J$-factor for looping) and to examine how the propensity for looping
depends upon the number of repressors, we needed looping data at a
number of different concentrations. At very low concentration, we
expect that there will be negligible looping because
neither of the operators will be bound by Lac repressor. At
intermediate concentrations, the equilibrium will be
dominated by states in which a single repressor tetramer is bound to
the DNA at the strong operator, punctuated by transient looping
events. In the very high concentration limit, each operator will be
occupied by a tetramer (see fig.~\ref{fig:thermomodel} below), making the formation of a loop nearly
impossible.

\begin{figure}
\begin{center}
\includegraphics[width=3.0in]{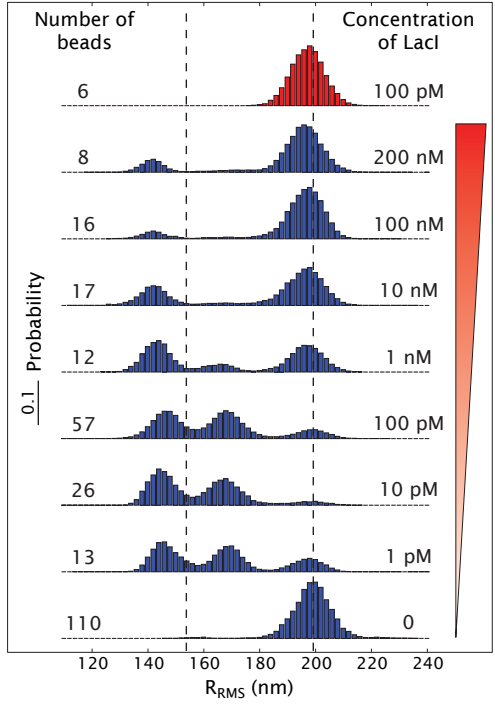}
\caption{\small{Concentration dependence of  the distribution of bead excursions.  The histograms
show the distribution of RMS motions averaged over 4~seconds at different concentrations of Lac repressor.
The blue histograms correspond to measurements for a length between operators of $\NLl=306$~bp (see fig.~\ref{fig:ConstructScheme}),
whereas the red histogram
is a control where $O1$ has been deleted.
The two dashed lines represent  the naively expected motion, based on our calibration measurements~\citep{calibration}. (See fig.~\ref{fig:PNCombined} for a more precise prediction of the peak locations.)
Representative single-molecule trajectories resulting in these histograms are shown in the Supplementary Material.}
} \label{fig:histogram}
\end{center}
\end{figure}
This progression of qualitative behavior is indeed seen in
fig.~\ref{fig:histogram}, which shows data from eight distinct
concentrations of Lac repressor, as well as a single-operator
control in which the DNA lacks a secondary operator. Throughout this
work we define {\it sequence length} or {\it loop length} as the
end-to-end distance between the operators as shown in
fig.~\ref{fig:ConstructScheme}. These curves correspond to a
sequence length of 306~bp and are generated by summing the
normalized histograms from {\it all} of the individual trajectories
for each concentration that pass our bead selection criteria (bead
selection criteria are discussed in detail in the Supplementary
Materials). A key feature of these data is the way in which the two
looped states are turned off as the concentration of Lac repressor
is increased to very high levels. This phenomenon is expected since
the Lac repressor exists always as tetramers under the conditions
used here{~\citep{Levandoski1996, Barry1999}}, and competition for
binding at the second operator between loose Lac repressor and Lac
repressor bound to the other operator is stronger as the
concentration of Lac repressor increases. However,  the two
different looped species have slightly different responses at high
repressor concentrations.  For example, at 1~nM concentrations, the
intermediate looped state has become very infrequent, whereas the
shortest looped state remains competitive. Similar concentration
dependence of Lac repressor mediated DNA looping was studied
previously~\citep{Vanzi2006} at 4~pM, 20~pM and 100~pM. Those
experiments revealed that  looping is suppressed as the
concentration goes up.

\begin{figure}
\begin{center}
\includegraphics[width=4in]{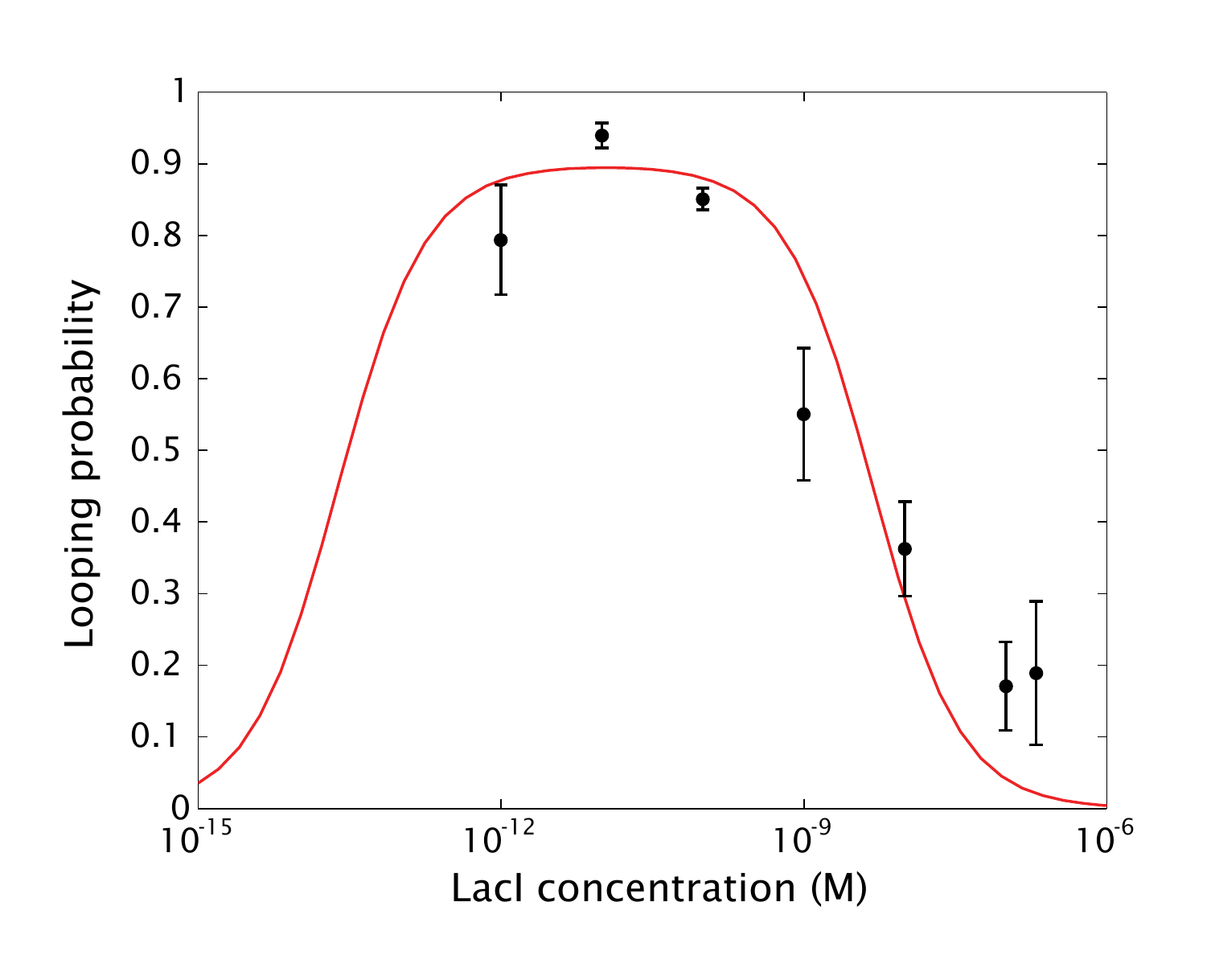}
\caption{\small{Looping probability $p_{\rm loop}$, at different concentration of Lac repressor. The DNA
used in these experiments  is 901~bp long and the loop length is $\NLl=306$~bp. The vertical axis gives
looping probability (fraction of time spent in either of the two looped states).
The fraction of time spent in the looped states was calculated for each bead individually and the mean and standard
error calculated for each construct.
The curve is a fit to the experimental data using the
statistical mechanics model described in the text.}
 }
\label{fig:loopingpro}
\end{center}
\end{figure}

One way to characterize the looping probability as a function of
concentration is shown in fig.~\ref{fig:loopingpro}.     There are
various ways to obtain data of the sort displayed in this  plot.
First, by examining the trajectories, we can simply compute the
fraction of time that the DNA spends in each of the different
states, with the looping probability given by the ratio of the time
spent in either of the looped states to the total elapsed time.  Of
course, to compute the time spent in each state, we have to make a
thresholding decision about when each transition has occurred. This
can be ambiguous, because trajectories sometimes undergo rapid jumps
back and forth between different states; it is not unequivocally
clear when an apparent transition is real, and when it is a random
fluctuation without change of looping state. A second way of
obtaining the looping probability is to use fig.~\ref{fig:histogram}
and to compute the areas under the different peaks and to use the
ratios of areas as a measure of looping probability. This method,
however, does not properly account for possible variation between
different beads, because they are all added up into one histogram. A
third alternative is to obtain the looping probability for each
\textit{individual} bead, by plotting its histogram and calculating
the area under that  subset of the histogram corresponding to the
looped states.   We used this last method  to calculate the mean
looping probability and the standard error for each construct, which
is shown in fig.~\ref{fig:loopingpro}.

These results can also be explored from a theoretical perspective
using the tools of statistical
mechanics~\citep{Bintu2005a,Bintu2005b,Saiz2005}.   The goal of a
statistical mechanical description of this system  is to compute the
probability of the various microstates available to the
repressor-DNA system as shown in fig.~\ref{fig:thermomodel}. The
simplest model posits 5 distinct
states~\citep{Wong2007,Finzi1995,Vanzi2006}: Both operators empty,
$Oid$ occupied by repressor without looping, $O1$ occupied by
repressor without looping, $Oid$ and $O1$ separately occupied by
single repressors and the looped state (the subtleties associated
with the statistical weight of the looped state are described in the
Supplementary Materials). The model does not take into account the
effect of non specific binding of Lac repressor to non-operator DNA,
because a simple estimate reveals that the vast majority of
repressors are free in solution rather than bound nonspecifically to
the tethered DNA.  We argue that this effect is negligible because
the equilibrium association constant of Lac repressor  to
non-operator DNA at conditions similar to ours is around
$10^6~\sim~10^7$
M$^{-1}$~\citep{deHaseth1977,OGorman1980,Revzin1977,Record1977,Kao-Huang1977,Wang1977,Barkley1981},
which is roughly six orders of magnitude less than the corresponding
quantity for specific binding
\citep{Zhang1993,Mossing1985,Hsieh1987,Horton1997,Goeddel1977,Falcon1999,Winter1981a}.
Given such association constants, the ratio between non specifically
bound Lac repressor and the free Lac repressor  in solution is given
as
\begin{eqnarray*}
\frac{[RD]}{[R]}&=& K_{NS} \times [D] \\
                         &\approx& 2 \times 10^{-5} ,
\end{eqnarray*}
where $[RD]$ is the concentration of non-specifically bound Lac
repressor, $[R]$ is the concentration of Lac repressor  in solution,
and $[D]$ is the DNA concentration, which is around 2~pM in our
experiment. For $[R] = 200$~nM, we have $[RD]\approx4$~pM, which is
far smaller than the concentration of Lac repressor in solution.

It is convenient to describe the probability of the various states
using both the language of microscopic binding energies (and looping
free energies) and  the language of equilibrium constants (and
$J$-factors). From a microscopic perspective, the key parameters
that show up in the model are the standard free energy changes for
repressor binding to the two operators, $\Delta \epsilon_{id}$ and
$\Delta \epsilon_{\mathrm{1}}$, the looping free energy $\Delta
F_{\mathrm{loop}}$ and the concentration of repressor $[R]$. The
binding energy here contains two components.  One is the standard
positional free energy required for bringing one Lac repressor
molecule to its DNA binding site at 1~M concentration of Lac
repressor.  The other is the rotational entropy loss times $-T$,
plus the interaction free energy due to the physical contact upon
protein binding~\citep{Bintu2005a,Bintu2005b,Saiz2006b}. The
associated free energy with each configuration gives the statistical
weights of the equilibrium probability (listed in the middle column
of fig.~\ref{fig:thermomodel}).  For example, to obtain the
probability of the looped state, we construct the ratio of state (v)
in the figure to the sum over all five states, as given by
\begin{eqnarray}\label{eq:ploopStatMech1}
    p_{\rm loop} &=& \left[8 {[R] \over 1~\mbox{M}} e^{-\beta(\Delta \varepsilon_1 +
                \Delta \varepsilon_{id} + \Delta F_{\rm loop})}\right] \\
             &&\left[1+ 4 {[R] \over 1~\mbox{M}} \left( e^{-\beta\Delta\varepsilon_1} +
                e^{-\beta\Delta\varepsilon_{id}}\right)+
                16 \left({[R] \over 1~\mbox{M}}\right)^2
                e^{-\beta(\Delta\varepsilon_1+\Delta \varepsilon_{id})} + \right. \nonumber \\
             &&\left. 8 {[R] \over 1~\mbox{M}} e^{-\beta(\Delta \varepsilon_1 +
                \Delta \varepsilon_{id} + \Delta F_{\rm loop})}\right]^{-1}, \nonumber
\end{eqnarray}
where $\beta=1/k_BT$. As detailed in the Supplementary Materials and
can be read off from the right column in fig.~\ref{fig:thermomodel},
this microscopic description is conveniently rewritten in terms of
the equilibrium constants and  $J$-factor for looping  as
\begin{equation}\label{eq:ploopThermo}
    p_{\rm loop} = \frac{{1 \over 2} \frac{[R]J_{\rm loop}}{K_1 K_{id}}}
                {1+\frac{[R]}{K_1} + \frac{[R]}{K_{id}} + \frac{[R]^2}{K_1 K_{id}}
                + {1 \over 2} \frac{[R]J_{\rm loop}}{K_1 K_{id}}}.
\end{equation}
Here $J_{\rm loop}$ is the average of the individual $J$ factors
corresponding to different loop topologies. These topologies can be
classified according to the orientation of each one of the operators
with respect to the binding heads. We define the state variables
$\alpha$ and $\beta$ that describe the orientation of $O1$ and
$Oid$, respectively, and that can adopt a value of either 1 or 2.
The average $J_{\rm loop}$ is then
\begin{equation}
    J_{\rm loop} = {1 \over 4} \sum_{\alpha,\beta} J_{\mathrm{loop},\alpha,\beta}.
\label{eq:Jtot}\end{equation}
An alternative to this scheme is to
construct the ratio $p_{\rm unloop}/p_{\rm loop}$. In the limit
where the strongest operator, $Oid$, is always occupied, this ratio
takes the simple, linear form
\begin{equation}\label{eq:pratioThermo1}
    p_{\mathrm{ratio}} = {2 K_1 \over J_{\rm loop}} + {2 [R] \over J_{\rm loop}}.
\end{equation}
This permits the determination of the  $J$-factor as the slope of a
linear fit of the form without necessarily a need to obtain $K_1$.
Below we discuss the validity of this particular model. For the
remaining data points at loop lengths other than 306~bp, where no
titration was done, we can use the relation
\begin{equation}\label{eq:pratio2J}
    J_{\rm loop}(L) = {p_{\mathrm{ratio}}(306~\rm{bp}) \over p_{\mathrm{ratio}}(L)} J_{\rm loop}(306~\rm{bp}).
\end{equation}
Just like in the titration case, this relation allows to obtain
$J_{\mathrm{loop}}$ without knowing $K_1$, as long as we know at
least one value of $J_{\mathrm{loop}}$ and its corresponding
$p_{\mathrm{ratio}}$.

\begin{figure}
\begin{center}
\includegraphics[width=4in]{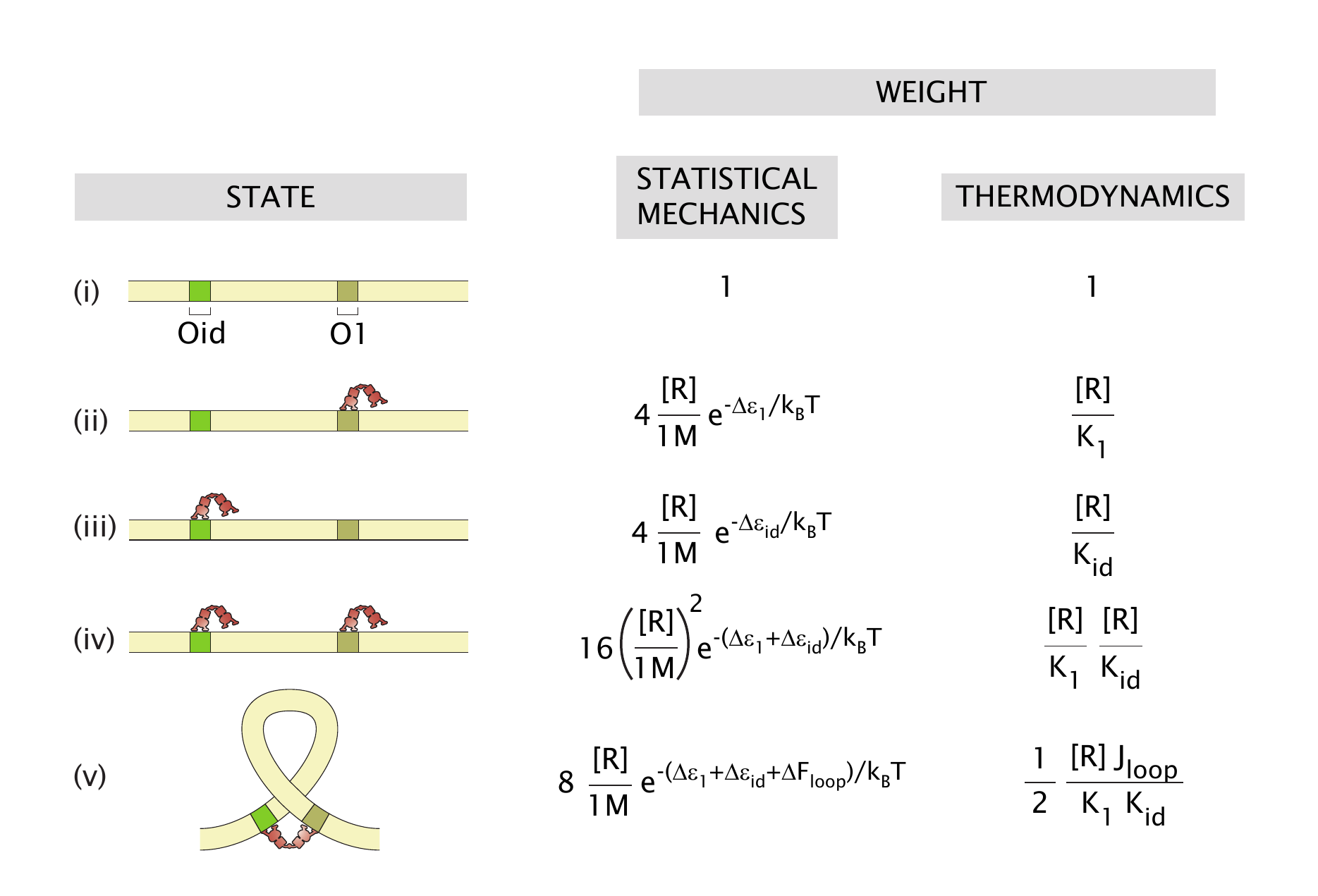}
\caption{\small{States and weights for the Lac repressor-DNA system~\citep{Bintu2005a}.  Each of the five state classes shown in the left column has a corresponding statistical weight given
by the product of the Boltzmann factor and the microscopic
degeneracy of the state.    All of the weights have been normalized by the weight of
the state in which
the DNA is unoccupied.  State (v)  is treated as a single looped state, even though there
are multiple distinct looped configurations.  The third column shows how to write these
statistical weights in the language of equilibrium constants and $J$-factors.  The derivation
of these weights and the relation between the statistical mechanical and thermodynamic
perspectives can be found in the Supplementary Materials.}
}
\label{fig:thermomodel}
\end{center}
\end{figure}

The data shown in fig.~\ref{fig:loopingpro} can be fit  in several
different ways as suggested by the three different formulae
characterizing the looping probability given above. The fit shown in
fig.~\ref{fig:loopingpro}  is a full nonlinear fit in which the
parameters $K_1$, $K_2$ and $J_{\rm loop}$ are treated as fitting
parameters.    Alternatively, using this same data of
fig.~\ref{fig:loopingpro}, we can actually obtain the looping free
energy, as well as the binding energies by fitting the data  to
eqn.~\ref{eq:ploopStatMech1}.  Finally, we can fit the data
corresponding to LacI concentrations of 10~pM and higher using the
linear model from eqn.~\ref{eq:pratioThermo1}. The results of these
different fits are shown in Table~\ref{tab:FitTable}. These results
are usefully contrasted with results of other experiments on the
{\it lac} operon, which are also summarized  in
Table~\ref{tab:FitTable}. We see from the table that the nonlinear
model fails to constrain the value of $K_{id}$ reliably. In the case
of the $O1$ binding constants we see a difference of almost two
orders of magnitude with published dissociation constants, which
translates into a difference of roughly  $4~k_{B}T$ in the binding
energy.

\begin{table}
  \centering
  \begin{tabular}{c|c|c|c}
        Parameter                   & Nonlinear fit         & Linear fit            & Literature value           \\  \hline \hline
        $J_{\rm loop}$                   & $8.6 \pm 6.3$~nM      & $52 \pm 40$~nM        & See fig.~\ref{fig:Jeff}   \\  \hline
        $\Delta F_{\rm loop}$           & $18.6 \pm 0.7~k_{B}T$ & $16.8 \pm 0.8~k_{B}T$ & N/A          \\  \hline \hline
        $K_{1}$                     & $0.49 \pm 0.45$~nM    & $3.0 \pm 2.5$~nM      & $10~\sim~22$~pM
            \citep{Zhang1993,Mossing1985,Hsieh1987,Horton1997,Goeddel1977,Falcon1999,Winter1981a} \\  \hline
        $\Delta \varepsilon_1$      & $-20.0 \pm 0.9~k_{B}T$& $-18.2 \pm 0.8~k_{B}T$& $-23.2~\sim~-24.0~k_{B}T$ \\  \hline \hline
        $K_{id}$                    & $0.2 \pm 2.3$~pM      & N/A                   & $2.4~\sim~8.3$~pM
            \citep{Frank1997} \\  \hline
        $\Delta \varepsilon_{id}$   & $-28 \pm 9~k_{B}T$    & N/A                   & $-24.1~\sim~-25.4~k_{B}T$ \\
  \end{tabular}
  \caption{\small{Results from the LacI titration experiments. The probability of looping as a
    function of Lac repressor concentration shown in fig.~\ref{fig:loopingpro} was fitted
    to the two non-linear models from eqns.~\ref{eq:ploopStatMech} and \ref{eq:ploopThermo}.
    A subset of the data corresponding to concentrations of LacI 10~pM and higher  is
    fitted to the linear model shown in eqn.~\ref{eq:pratioThermo1} and its
    statistical mechanics counterpart. The literature values correspond
    to bulk binding assays performed in concentration ranges close to our TPM buffer
    conditions.}
    }\label{tab:FitTable}
\end{table}

One of the challenges of single-molecule experiments like those
described here is that the concentration of protein introduced into
the system may not correspond to the actual concentration ``seen''
by the DNA that is tethered to the surface. For example, some of the
protein might be lost as a result of nonspecific binding to the
microscope cover slip. From the linear model shown in
eqn.~\ref{eq:pratioThermo1} it follows that any error in the
concentration will translate linearly into an error in $J_{\rm
loop}$ and $K_{1}$. Therefore, in order for the above discrepancy to
be explained solely by surface effects on the LacI concentration we
would have to have a difference of between one and two orders of
magnitude between the concentration of the stock that  flowed into
the chamber and the actual free concentration inside of it.

\begin{figure}
\begin{center}
\includegraphics[width=4 in]{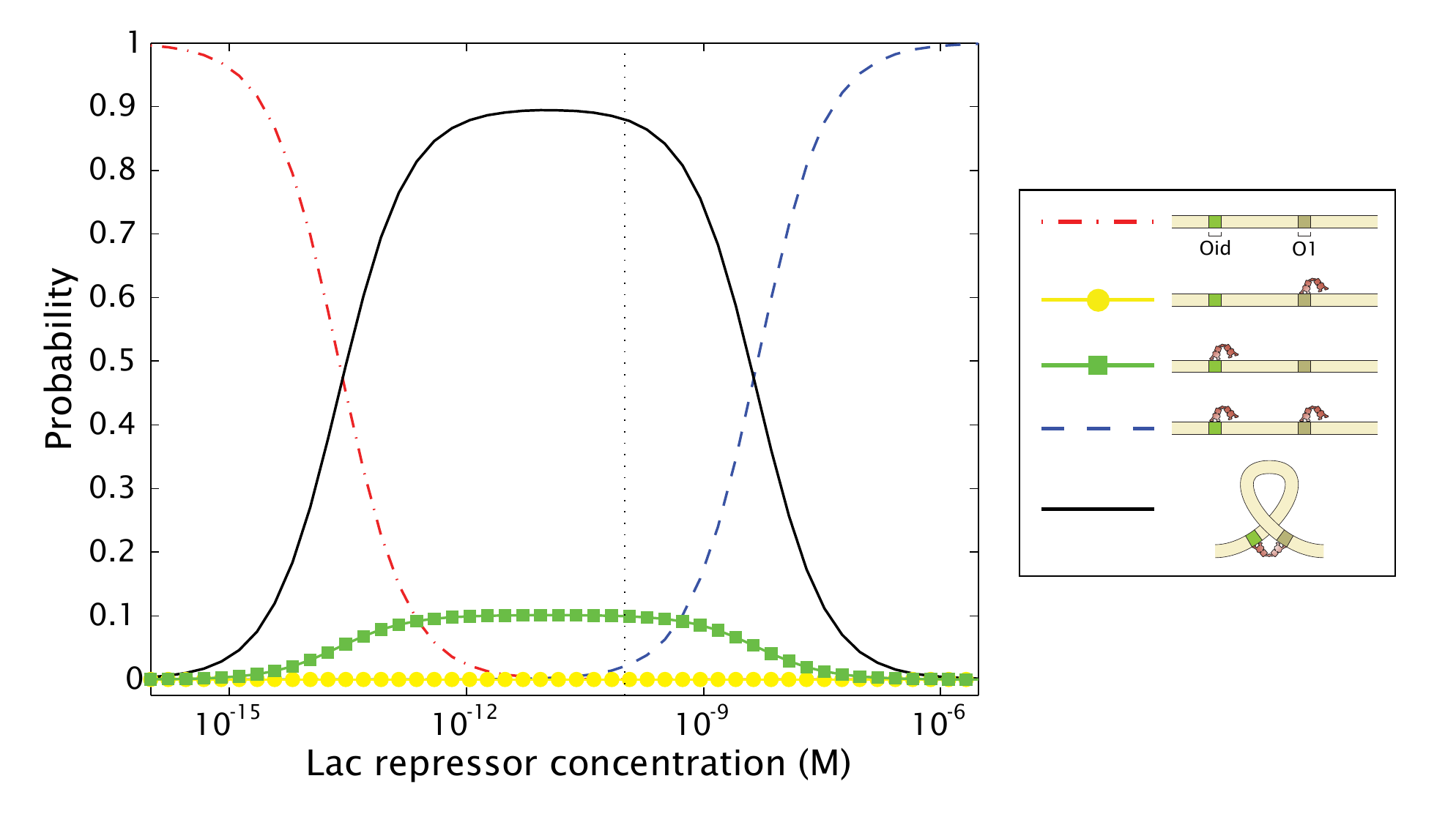}
\caption{\small{Probabilities  for different states of Lac repressor and operator DNA.
The curves  show the probabilities of the five classes of microscopic states
used in the statistical mechanics model based upon  parameters
shown in table \ref{tab:FitTable}.  The vertical line corresponds to the concentration
at which the loop length experiments in the remainder of the paper are performed.}
}
\label{fig:modelpro}
\end{center}
\end{figure}

Once the parameters that characterize the model are in hand, we can
plot the probability of all five possible states as a function of
the Lac repressor concentration as shown in fig.~\ref{fig:modelpro}.
This  figure reveals  that at the concentrations we normally use
($[R]=100$~pM), the system is dominated by the looped state and the
state with single occupancy of $Oid$. A detailed discussion of the
significance of the looping free energies (or the $J$-factors) will
follow later in the paper once we have explored the question of the
length dependence of DNA looping in the {\it lac} operon.

\subsection{Length dependence}

\subsubsection{1bp resolution for a whole helical turn: $\NLl=~$300~bp to 310~bp}

\begin{figure}
\begin{center}
\includegraphics[width=5.2in]{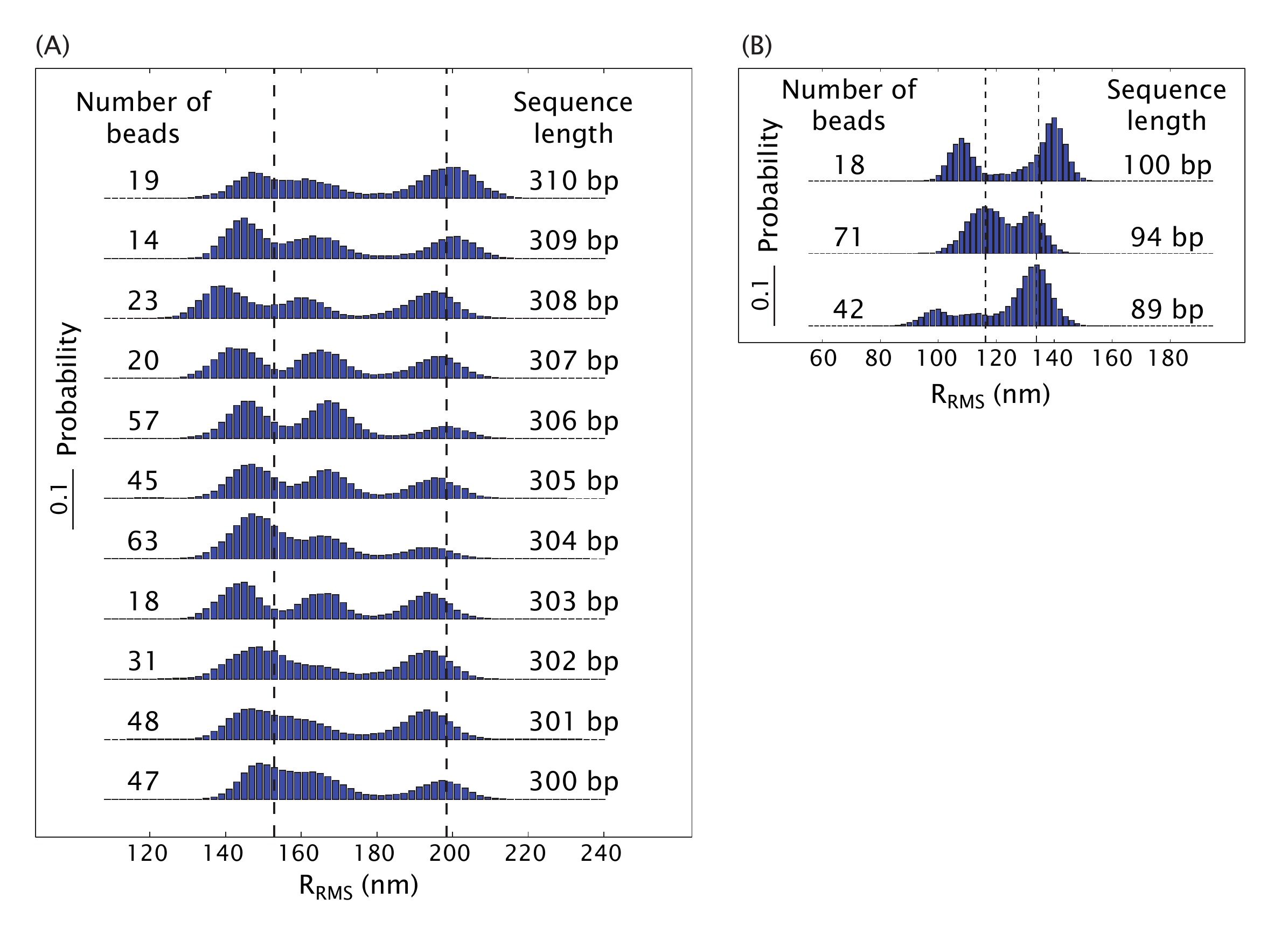}
\caption{\small{Length dependence of DNA looping.  (A) Histogram of the tethered Brownian motion for DNAs with two Lac repressor binding sites spaced from $\NLl=300$~bp
(bottom) to 310~bp (top).  (B) Histogram of the
Brownian motion for DNAs with two Lac repressor binding sites spaced at $\NLl=89$, 94 and 100~bp.
The two dashed lines represent the naively expected motion based on our calibration measurements for
the full length tether and the same DNA when the center to center distance between operators is subtracted from
the tether length. (Again see also fig.~\ref{fig:PNCombined}.)}  Representative
traces for each of the lengths shown here can be found in the Supplementary Material.
}
\label{fig:histogrambp}
\end{center}
\end{figure}

The beautiful {\it in vivo} repression experiments of
\cite{Muller1996} demonstrate that the length of the DNA loop formed
by Lac repressor strongly affects the probability of loop formation
(especially for loop lengths less than 150~bp). In particular, those
authors (and others) \citep{Lee1989,Law1993,Becker2005,Becker2007}
have observed ``phasing'': The relative orientations of the two
operators changes the ease with which repressor can bind.  Similar
phasing effects have been observed in {\it in vitro} cyclization
assays \citep{Shore1983a, Cloutier2004,Cloutier2005, Du2005}.  What
has not been clear is how to concretely and quantitatively relate
these results on DNA mechanics from the {\it in vivo} and {\it in
vitro} settings and how to use such insights to better understand
the interplay between the physical and informational properties of
DNA {\it in vivo}. Our idea was to systematically examine the same
progression of DNA lengths that have been observed {\it in vivo},
but now using TPM experiments. To that end, we have measured TPM
trajectories for a series of interoperator spacings measured in 1~bp
increments. The results of this systematic series of measurements
for DNAs harboring operators spaced over the range
$\NLl=300~\sim~310$~bp  are shown in fig.~\ref{fig:histogrambp} (as
are the results for several shorter lengths to be discussed in the
next section). Each plot shows the probability of the three states
for a particular interoperator spacing.

\begin{figure}
\begin{center}
\includegraphics[width=5.0 in]{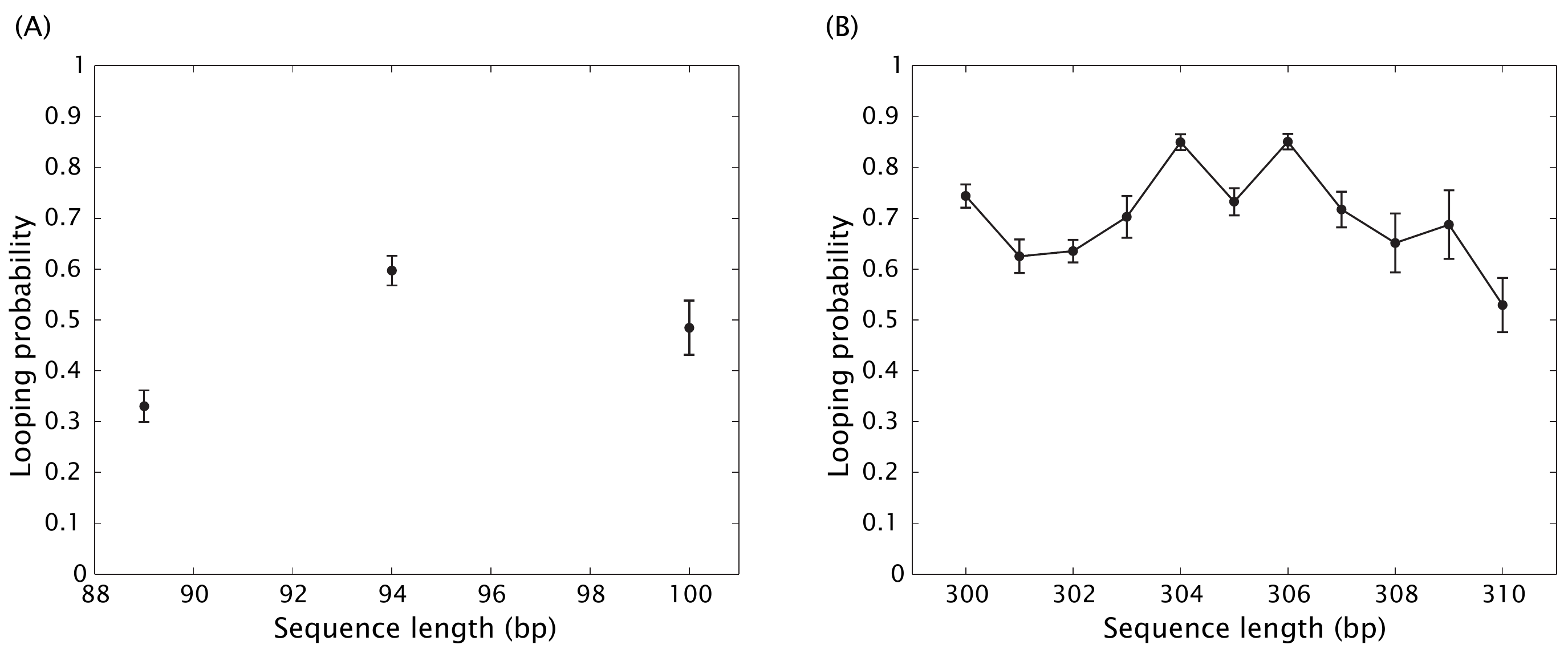}
\caption{\small{Looping probability $p_{\rm loop}$, as a function of interoperator spacing.  (A) Looping probability
for short constructs.  (B) Looping probability for one full helical
repeat. These probabilities are obtained by averaging over the $p_{\mbox{loop}}$ of each bead. The error bars
correspond to the standard error associated with this magnitude. For more information see Supplementary Materials.}
}
\label{fig:lp}
\end{center}
\end{figure}

The data can be converted into a plot of the dependence of the
looping probability on interoperator spacing as shown in
fig.~\ref{fig:lp}. This figure shows $p_{\rm loop}$  as a function
of the DNA length between the two operators. The looping probability
shows a weak dependence on the interoperator spacing but reveals no
conclusive signature of phasing; to really detect such phasing with
confidence, however, would require more measurements in single
basepair increments. The maximum looping is achieved when the  two
binding sites are 306~bp apart, suggesting that at this distance,
the two sites are  in an optimal phasing orientation for binding of
the two heads of Lac repressor.  The ability to form stable
out-of-phase (two binding sites are on the opposite side of the DNA)
loops with only a small reduction in stability is consistent with
previous studies~\citep{Wong2007}. The relatively stable looping
over the entire helical repeat is also consistent with  the
relatively constant repression level {\it in vivo}  for similar
interoperator spacing \citep{Muller1996}.

\begin{figure}
\begin{center}
\includegraphics[width=5.0in]{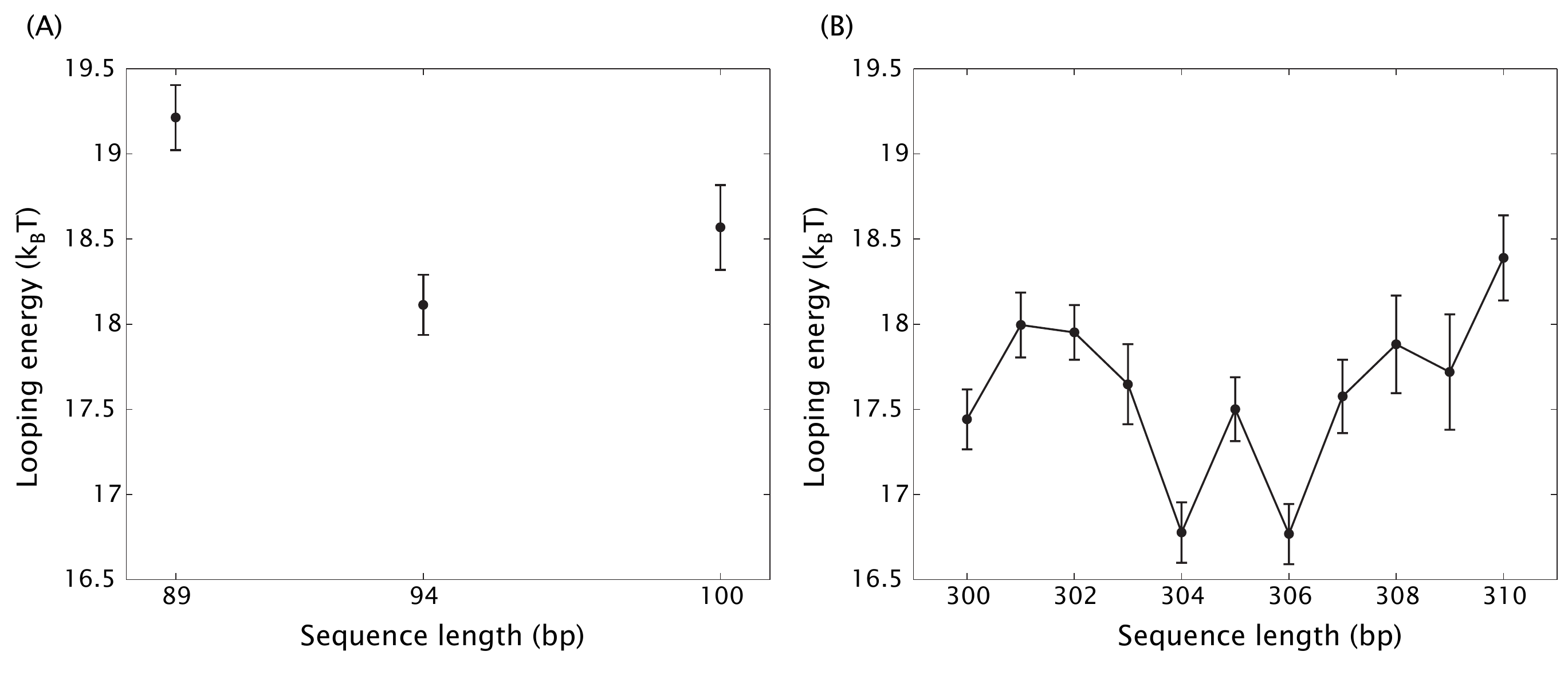}
\caption{\small{Length dependence of free energy of looping, defined via eqn.~\ref{eq:ploopStatMech1} with choice of reference concentration 1~M. (A) Looping free energy
for short constructs.  (B) Looping free energy for a full helical repeat.}}
\label{fig:FloopLength}
\end{center}
\end{figure}

\begin{figure}
\begin{center}
\includegraphics[width=4.2in]{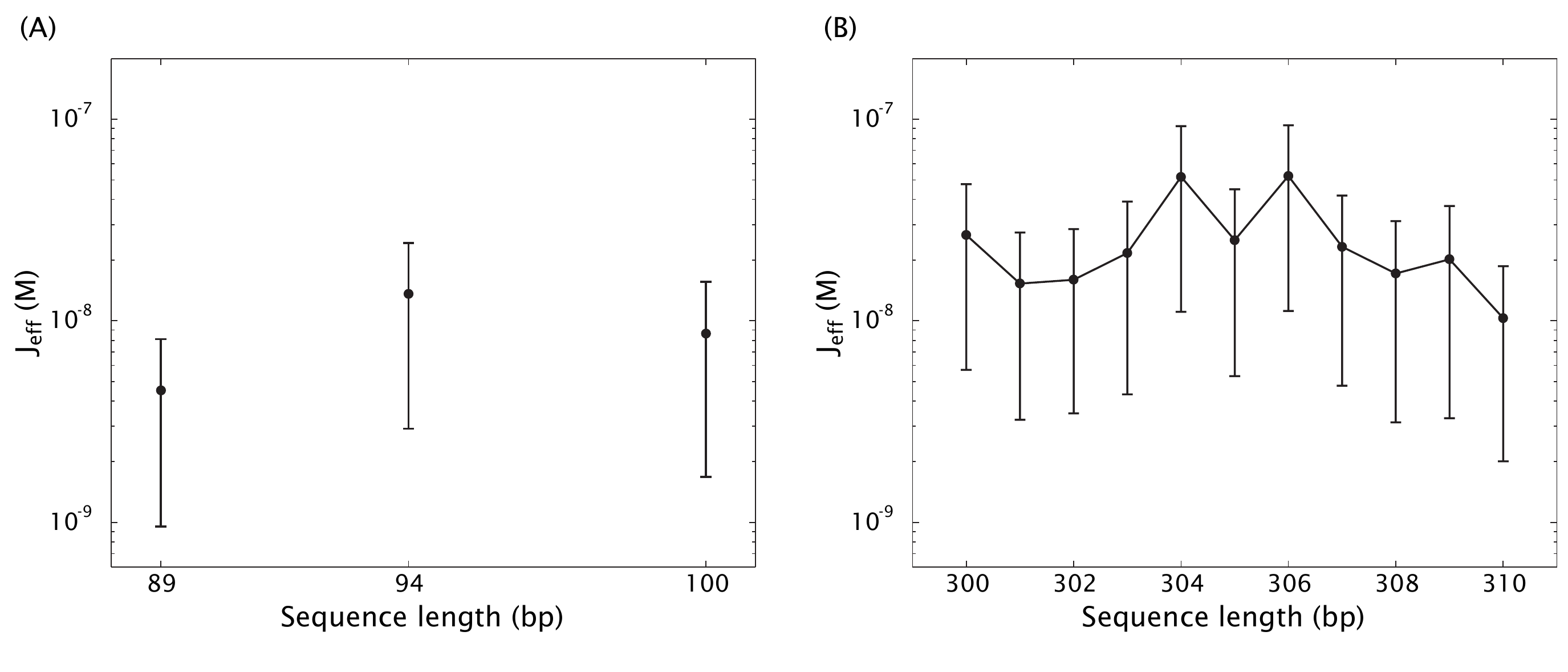}
\caption{\small{Looping $J$-factor resulting from TPM measurements.  (A) Effective $J$-factor
for looping resulting from TPM data on short constructs.  (B) Effective $J$-factor
for looping resulting from TPM data on a full helical repeat.}
}
\label{fig:Jloop}
\end{center}
\end{figure}

As already indicated in Table~\ref{tab:FitTable}, the looping
probability can be converted into  a corresponding looping free
energy based on the statistical mechanics  model described above and
culminating in eqn.~\ref{eq:ploopStatMech1}.    The results of such
calculation are shown in fig.~\ref{fig:FloopLength}. The
measurements on length dependence permit us to go beyond the
concentration dependence measurements by systematically exploring
how the phasing of the two operators impacts the free energy of DNA
looping.   One might expect that when the two operators are on
opposite sides of the DNA, additional twist deformation energy is
required to bring the operators into good registry for Lac repressor
binding. Our results show that the phasing effect imposes an energy
penalty $\Delta F_\mathrm{loop}$ that differs by only about 1.5
$k_\mathrm{B}T$ between the in-phase and out of phase cases. An
alternative interpretation of these same results on looping
probability is offered by the $J$-factor for looping as shown in
fig.~\ref{fig:Jloop}.

To get a feel for the energy scale associated with twist
deformations, we perform a simple estimate. Twisting DNA for a
torsional angle $\theta$ requires energy
\begin{eqnarray}
\Delta F_{t} = k_BT\xi_{tp}\theta^2/2L
\end{eqnarray}
where $\xi_{tp}$ is the torsional persistence length for double
stranded DNA,  which is around 250
bp~\citep{Record1981,Strick2000,Moroz1998}. $L$ is the DNA length.
For half a helical turn twist,  $\theta = \pi$ and $L=300$~bp. The
energy introduced for half a helical turn is around $4.11~k_BT$. Our
experimentally determined looping energy difference between in-phase
and out-of-phase DNA, about $1.5~k_BT$, is indeed comparable in
magnitude to this estimate. Our simple estimate is somewhat high, in
part because it neglects the fact that in addition to twisting, a
loop can writhe to accommodate a nonideal operator phasing.
Additionally, the observed small magnitude of our observed phasing
modulation may reflect partially canceling out-of-phase
contributions of different topologies~\citep{Towles2008}, not a low
free energy cost for twisting. Finally,  the Lac repressor itself is
flexible, and so can partially compensate for nonideal phasing.

\subsubsection{Sub-persistence length loops}

One of the intriguing facts about the architecture of regulatory
motifs that involve DNA looping is that often the loops formed in
these systems have DNA lengths that are considerably shorter than
the persistence length of DNA (i.e. 150~bp).  For example, in the
{\it lac} operon, one of the two wild-type loops has a length of
92~bp. However, this trend goes well beyond the {\it lac} operon as
is seen for a variety of different architectures found in {\it E.
coli}, for example \citep{Garcia2007a}.  As a result, it is of great
interest to understand the interplay between transcriptional
regulation and corresponding mechanical manipulations of DNA this
implies.

So far, we have considered loops that are roughly two-fold larger
than the persistence length through our investigation of one full
helical repeat between 300 and 310~bp. To begin to develop intuition
for the mechanism of loop formation in the extremely short loops
exhibited in many regulatory architectures, we have examined three
different lengths: 89, 94 and 100~bp.   One of the reasons that the
examination of these loops is especially important is that it has
been speculated that the {\it in vivo} formation of these loops
either requires special supercoiling  of the DNA or the assistance
of helper proteins that prebend the DNA \citep{Garcia2007a}.
However, as indicated by the TPM results shown in
fig.~\ref{fig:histogrambp}(B), even in our controlled {\it in vitro}
setting, where neither of these mechanisms can act, Lac repressor is
nevertheless able to form DNA loops.  The essence of these
experiments is identical to those described earlier in the paper
except that now the overall tether lengths are shorter so as to
ensure that the loops are detectable. (Representative TPM
trajectories for these lengths are shown in the Supplementary
Material.)
  It is clear from the histograms that of the three lengths
we have investigated, loop formation is most favorable at 94~bp.  Interestingly,
it also appears that different loops are being formed for the in-phase
and out-of-phase cases as evidenced by the changes of relative strengths among the looping peaks for
the different constructs.  The looping free energy and $J$-factor
for looping for these short constructs are shown in
figs.~\ref{fig:FloopLength}(A) and~\ref{fig:Jloop}(A).

\subsection{Analysis of the TPM Experiment}\label{sec:AnalysisTPM}
Both the observed length and sequence dependence of the formation
of a repression complex are intriguing from the perspective
of DNA mechanics.    In particular, DNA is not a passive mechanical
bystander in the process of transcriptional regulation.
 To better understand the experiments carried out here
 and how they might shed light on the interplay of transcription
 factors and their target DNA, we have
appealed to two classes of models: i) statistical mechanics models
of the probability of DNA-repressor complex formation  which depends
upon the looping free energy (these models were invoked earlier in
the paper to determine the looping free energy) and ii) Monte Carlo
simulations of the TPM experiment itself which include the
energetics of the bent DNA and excluded volume interactions of the
bead with the cover slip. Our Monte Carlo calculations allow us to
compute how easily loops form, based on a mathematical model of DNA
elasticity. For illustration, we have chosen a linear-elasticity
model, that is, a model in the class containing the wormlike chain,
but any other elastic theory of interest can be used with the same
calculation strategy. Details of these calculations appear in
\citep{Towles2008}.

One of the puzzles that has so far been unresolved concerning DNA
mechanics at short scales is whether  {\it in vivo} and {\it in
vitro} experiments tell a different story.  In particular, {\it in
vivo}  experiments, in which repression of a given gene is measured
as a function of the interoperator spacing
\citep{Muller1996,Becker2005}, have the provocative feature that the
maximum in repression (or equivalently the minimum in looping free
energy) correspond to interoperator spacings that are shorter than
the persistence length.  Some speculate that this {\it in vivo}
behavior results from the binding of helper proteins such as the
architectural proteins HU, H-NS or IHF
\citep{Becker2005,Becker2007,Garcia2007a} or the control of DNA
topology through the accumulation of twist. In the TPM measurements
reported here, there are neither architectural proteins nor proteins
that control the twist of the DNA.   As a result, these experimental
results serve as a jumping off point for a quantitative
investigation of whether DNA at length scales shorter than the
persistence length behaves more flexibly than expected on the basis
of the wormlike chain model.   To address this question, we
performed a series of simulations of the probability of DNA looping
for short, tethered DNAs like those described here using,  a variant
of the wormlike chain model to investigate the looping probability.
Our theoretical model used \textit{no fitting parameters;} the few
parameters defining the model were obtained from other, non-TPM,
experiments.

The fraction of time spent in the looped configuration is controlled
by several competing effects. For example, suppose that a repressor
tetramer is bound to the stronger operator, $Oid$. Shortening the
interoperator spacing reduces the volume over which the other
operator ($O1$) wanders relative to the second binding site on the
repressor, increases the apparent local ``concentration'' of free
operator in the neighborhood of that binding site, and hence
enhances looping. But decreasing the interoperator spacing also has
the opposite effect of discouraging looping, due to the larger
elastic energy cost of forming a shorter loop. Moreover, a shorter
overall DNA construct increases the entropic force exerted by
bead--wall avoidance, again discouraging looping~\citep{sega06a}. To
see what our measurement of this looping equilibrium tells us, we
therefore needed to calculate in some detail the expected local
concentration of operator (the ``looping $J$ factor'') based on a
particular mathematical model of DNA elasticity. We chose a
harmonic-elasticity model (a generalization of the traditional
wormlike chain model), to see if it could adequately explain our
results, or if, on the contrary, some non-harmonic model (for
example the one proposed in~\citep{yan04a,wigg05a}) might be
indicated.

To perform the required calculation, we modified the Gaussian
sampling method previously used in
\citep{sega06a,Nelson2006,czap06a,nels07a} (see \sref{mcc} and
\cite{Towles2008}). Our code generated many simulated DNA chains,
applied steric constraints~\citep{sega06a}, and reported what
fraction of accepted chain/bead configurations had the two operator
sites at the correct relative position and orientation for binding
to the tetramer, which was assumed to be rigidly fixed in the form
seen in PDB structure 1LBG~\citep{Lewis1996}. Once this fraction has
been computed, it is straightforward to relate it to the looping $J$
factor \cite{Towles2008}. To generate the simulated chains, we
assumed a linear (harmonic, or wormlike-chain type) elastic energy
function at the junctions in a chain of finite elements. Our energy
function accounted for the bend anisotropy and bend--roll coupling
of DNA, and yielded a value for the persistence length
$\xi=44\,\nmunit$ appropriate for our experiment's buffer
conditions~\citep{stri98a,wang97a}. Our model did not account for
sequence dependence, but this simplification is appropriate for
comparison to our experimental results, which used random-sequence
DNA. The simulation treated the bead and the microscope slide as
hard walls and accounted for bead--wall, bead--chain, and
wall--chain avoidance; we did not consider any interactions
involving the repressor tetramer other than binding.

\begin{figure}
\begin{center}
\includegraphics[width=5in]{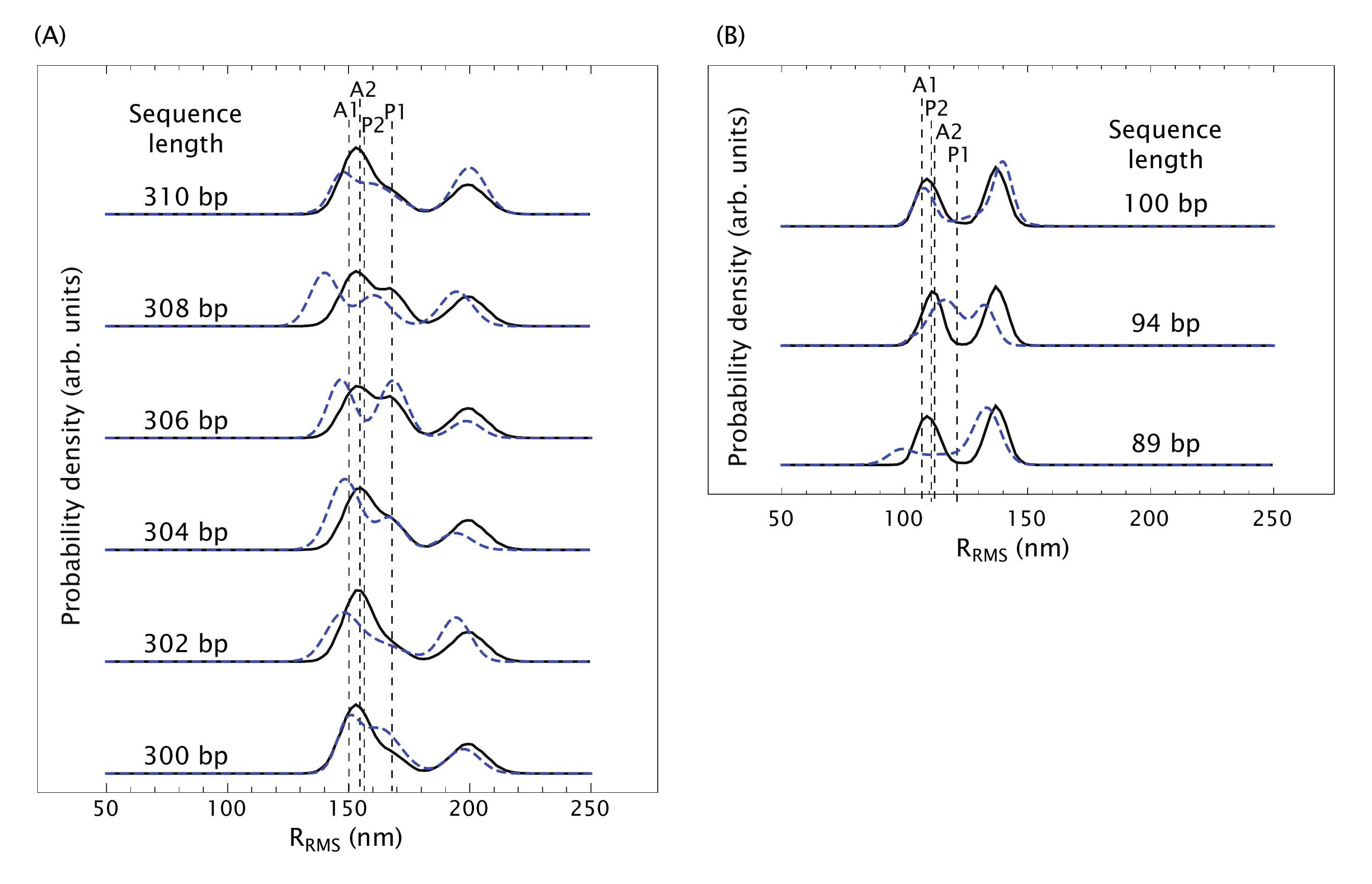}
\caption{\small {Theory and experiment for the probability density functions of RMS bead excursion for (A)
our six ``long chain'' constructs and (B) our three ``short chain'' constructs. \textit{Blue dashed curves}
show the data in fig.~\ref{fig:histogrambp}, represented as sums of three Gaussians. \textit{Black curves} show our theoretically predicted distributions. Because our simulation results were not fits
to the data, they did not reproduce perfectly the ratio of looped to unlooped occupancies. For visualization, therefore, we have adjusted this overall ratio
by a factor common to all six curves. This rescaling does not affect the locations of the peaks, the relative weights of the two looped-state peaks, nor the dependences of weights on loop length
$\NLl$, all of which are zero-fit-parameter predictions of our model.
The model yields these histograms as the sum of five contributions, corresponding to the four looped
topologies and the unlooped state. The separate RMS displacements for each individual loop
topology, for the $310\,$~bp case in (A) and for the $100\,$~bp case in (B), are also shown, labeled
according to the scheme in \citep{Swigon2006}.}}
\label{fig:PNCombined}
\end{center}
\end{figure}

The symmetry of each LacI dimer implies four energetically
equivalent ways for the two operators to bind when forming a loop,
and hence four topologically distinct loop configuration
classes~\citep{Zhang2006, Swigon2006, Geanacopoulos2001,
Balaeff2006,Towles2008}. We first asked whether this multiplicity of
looped states could explain the general structure of the excursion
distributions seen in fig.~\ref{fig:histogrambp}. Accordingly, we
made histograms of the distance between wall attachment point and
bead center for our simulated chains. Fig.~\ref{fig:PNCombined}
shows a subset of the same experimental data seen in
fig.~\ref{fig:histogrambp}, together with the simulation results.
Although the correspondence is not perfect, it is clear that the
simple physical model of looping outlined above can account for many
features of the data, for example the locations of the looped peaks
and their relative strengths, including the variation as loop length
is changed. We acknowledge that we have no definitive reply to the
argument that the apparent direct transitions between the B and M
peaks of our distributions seem to require an open-to-closed
conformational switch in the tetramer \citep{Wong2007}. We merely
point out that the existence of three peaks in the distribution,
with the the observed locations, is not by itself conclusive
evidence  of such a switch. (Indeed, Villa {\it et al.}\ have argued
that the opening transition does not occur \citep{Villa2005}.)

We were also interested to see if the high incidence of looping
observed in our experiments on short (sub persistence length) loops
was compatible with the hypotheses above, or if on the contrary it
demanded some modification to those hypotheses. Accordingly, we
asked the simulation to compute the average $J$ factor for loop
lengths near 305~bp, and also for loop lengths near 95~bp. As
discussed in ref.~\citep{Towles2008}, the result of the simulation
was that the ratio of these quantities is $\bar J_{\rm
loop}(95\,\bpunit)/\bar J_{\rm loop}(305\,\bpunit)\approx 0.02$. In
contrast, fig.~\ref{fig:Jloop} shows that the experimental ratio is
$\approx 0.35\pm 0.1$, roughly 20-fold larger than the theoretical
value. Our experimental results and those of our MC calculations for
$\bar J_{\rm loop}$ as a function of loop length are shown in
fig.~\ref{fig:Jeff}.

We conclude that the hypotheses of linear elasticity,  a rigid
protein coupler, and no nonspecific DNA--repressor interactions,
cannot explain the high looping incidence seen in our experiments.
(Special DNA sequences loop even more easily than the random
sequences reported here.) One possible explanation, for which other
support has been growing, is the hypothesis of DNA elastic breakdown
at high curvature \citep{yan04a,wigg05a,wigg06b}. An alternative
hypothesis is that for our shorter loops, {\it both} the lower and
the intermediate peaks in our distributions of bead excursion
correspond to the some alternative, ``open'' conformation of the
repressor tetramer
\citep{Ruben1997,Mehta1999,Edelman2003,Morgan2005,Zhang2006,Zhang2006b,Swigon2006}.
To be successful, however, this hypothesis would have to pass the
same quantitative hurdles to which we subjected our hypotheses. It
would be very useful for future TPM experiments to instead examine
other DNA-binding proteins known to be less flexible than LacI.

\section{Discussion}
The regulatory regions on DNA can often be as large as (or even larger than) the genes they
control.     The relation between the biological mechanisms of
transcriptional control and the physical constraints put
on these mechanisms as a result of the mechanical properties of
the DNA remains unclear.     One avenue for
clarifying action at a distance by transcription factors is systematic
single-molecule experiments, which probe the dynamics of loop
formation for different DNA architectures (i.e.\ different sequences, different
transcription factor binding strengths, different distances between
transcription factor binding sites) to complement systematic
{\it in vivo} experiments that explore these
same parameters.  In this paper, we have described an example of
such a systematic
series of measurements, which begins to examine how the formation
of transcription factor-DNA complexes depend upon parameters such
as transcription factor concentration and the length of the DNA implicated
in the complex.

\begin{figure}
\begin{center}
\includegraphics[width=6.0truein]{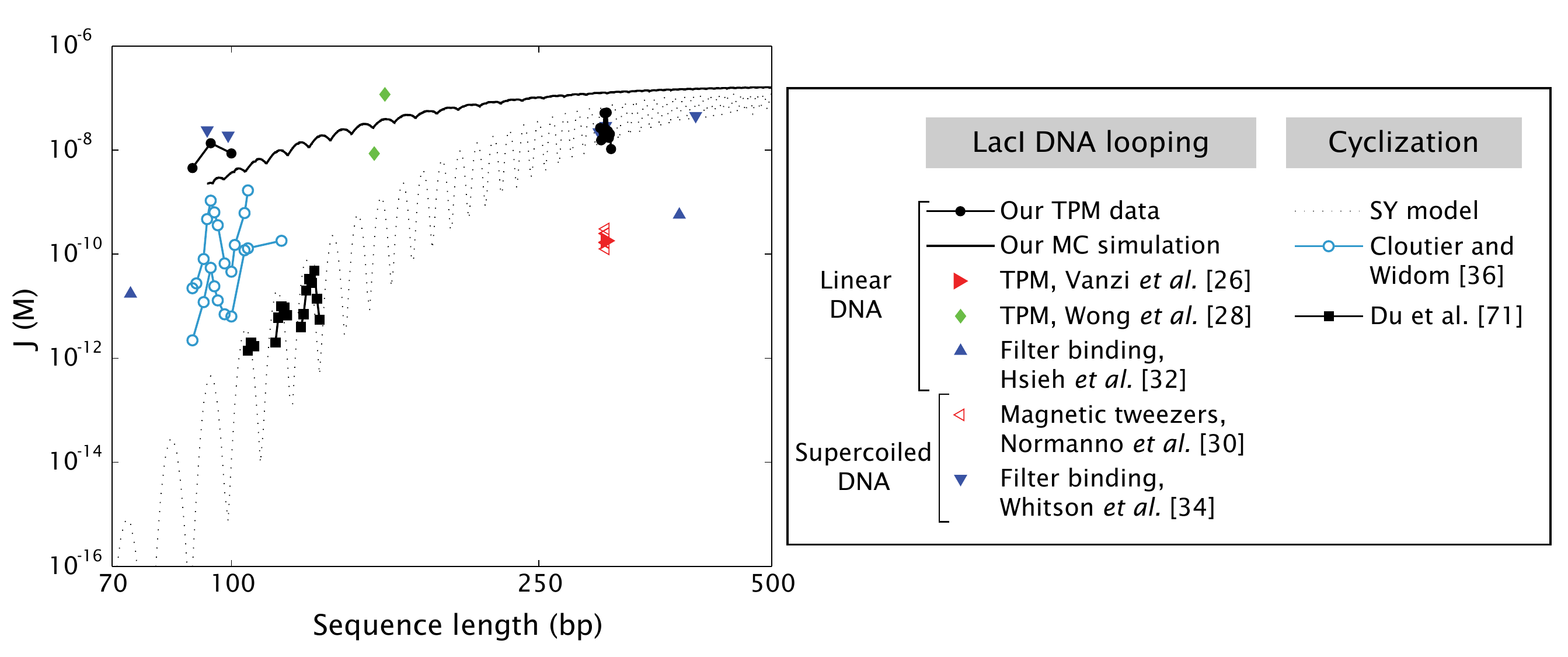}
\caption{\small{Effective $J$-factor from different experiments.
Although the $J$ factor obtained from cyclization experiments is not directly comparable
to the looping $J$ factor studied in this
paper (due to the differences in geometry), we  present the two quantities together
as functions of loop length to summarize the work from many groups'.
Error bars have been omitted for clarity. The filter binding data is an order of magnitude estimate.}}
\label{fig:Jeff}
\end{center}
\end{figure}

In the case of the {\it lac} operon, our {\it in vitro} measurements
demonstrate that the formation of the repressor-DNA complex does not
require any helper proteins, nor does it call for supercoiling of the
DNA (as appears to be required in other bacterial regulatory
architectures \citep{Schleif1992,Matthews1992}). Further, we find
that even in the absence of these mechanisms, which can only enhance
the probability of loop formation, the formation of DNA loops by Lac
repressor occurs more easily than would be expected on the basis of
traditional views of DNA elasticity.  A summary of the various
measurements of short-length DNA cyclization and looping is shown in
fig.~\ref{fig:Jeff}.    The idea of this figure is to present the
diversity of data that weighs in on the subject of short length DNA
elasticity.  In particular, several sets of controversial
measurements on DNA cyclization present different conclusions on the
ease of this process at lengths of roughly 100~bp.  TPM
experiments like those presented here offer another avenue to resolve this issue, one that does not involve the complex ligase enzyme, the need to ensure a specific kinetic regime, nor other
subtleties of the ligation reaction inherent in cyclization measurements.   However,
as seen in the figure, even here there are unexplained discrepancies between
different TPM experiments which call for continued investigation.

Several intriguing mysteries remain which demand both further
experimentation as well as theoretical analysis, e.g.: i)~why are
the probabilities of DNA loop formation systematically higher than
would be expected on the basis of traditional arguments about DNA
elasticity, and ii)~what is the significance of three repressor
binding sites in the wild-type {\it lac} operon?  To explore these
questions, TPM experiments with different DNA sequences between the
two operators, as well as with Lac repressor mutants that are less
flexible, would go a long way towards clarifying the mechanisms at
work and would provide a basis for examining the even richer action
at a distance revealed in the eukaryotic setting.

\section{Materials and Methods}
\label{section:MM}
\subsection{Plasmid DNAs}
\begin{figure}
\begin{center}
\includegraphics[width=6in]{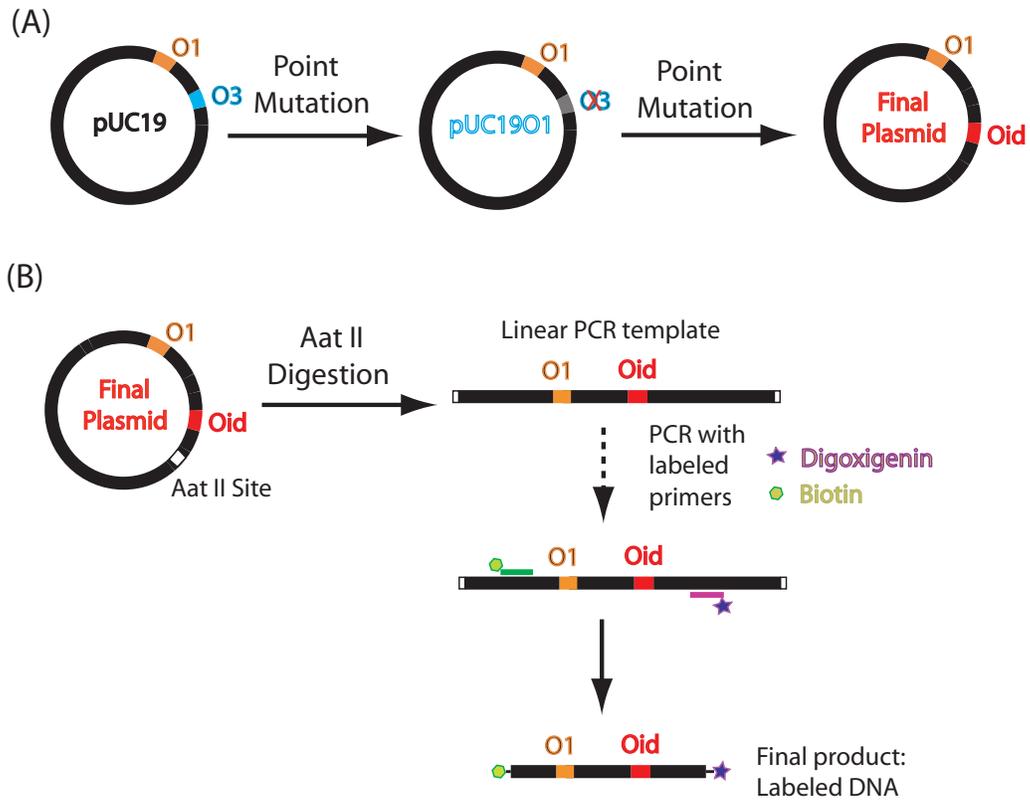}
\caption{\small{Synthesis of DNA construct. (A) Schematic of the
procedure  for construction of the plasmid with two Lac repressor binding sites.  (B) Schematic
of the protocol for producing labeled DNA using a PCR reaction with
labeled primers. }}
\label{fig:materials}
\end{center}
\end{figure}

Plasmid DNAs, bearing two Lac repressor binding sites spaced at a
designed distance, are created using a point mutation method
(QuikChange site-directed mutagenesis, Stratagene) on plasmid pUC19.
Plasmid pUC19 was chosen as a starting template because it is not
only a high copy plasmid but also contains two Lac repressor binding
sites: $O1$ and $O3$.  The procedure for creating two binding sites
separated by the desired distance  from template pUC19 is
illustrated in fig.~\ref{fig:materials} (a).  We first mutate six
basepairs in the $O3$ site converting it to $O3^*$ in a way that
eliminates the binding affinity for this site \citep{Oehler1994}.
The resulting plasmid is called pUC19O1 indicating it only has a
single $O1$ site.  To construct another binding site on the pUC19O1
plasmid,  we replace 20bp with the Lac repressor binding sequence
$Oid$ at a series of locations differing  by 1bp increments in their
distance from $O1$ using the mutagenesis method again.  For some of
the secondary site construction,  we have to use either deletion or
addition from already made plasmids with two designed binding sites.
The details on primers and templates used in this process are listed
in Table \ref{tab:mutprimers}.  The final product contains two
binding sites $O1$ and $Oid$ spaced at the desired distance.

The short loop DNA (89, 94 and 100~bp) was constructed in the
following way. Plasmid pZS22-YFP was kindly provided by Michael
Elowitz. The main features of the pZ plasmids are located between
unique restriction sites \citep{Lutz1997}. The YFP gene comes from
plasmid pDH5 (University of Washington Yeast Resource Center
\citep{Rosenfeld2005}).

\begin{figure}
\begin{center}
\includegraphics[width=5in]{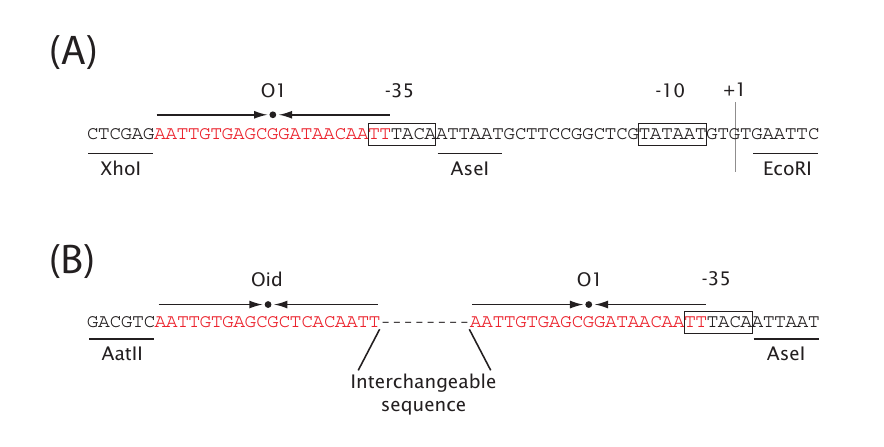}
\caption{\small{Promoter regions of the different short loop constructs. {\bf (A)}
Promoter region of pZS25-YFP which has a variant of the lacUV5
promoter and an $O1$ operator upstream overlapping the -35 region.
{\bf (B)} Final construct that allows to insert arbitrary DNA
sequences between a $Oid$ and $O1$ operators.}} \label{fig:lacUV5O1-45}
\end{center}
\end{figure}

A variant of the lacUV5 promoter \citep{Muller-Hill1996} was
synthesized and placed between the EcoRI and XhoI sites of pZS22-YFP
in order to create pZS25'-YFP. This promoter included the -35 and
-10 regions of the lacUV5 promoter, an AseI site between the two
signals and a $O1$ operator at position -45 from the transcription
start as shown in fig.~\ref{fig:lacUV5O1-45}(A).

The random sequence E8-89 \citep{Cloutier2004,Cloutier2005} was
obtained by PCR from a plasmid kindly provided by Jonathan Widom.
The primers used had a flanking AatII site and $Oid$ operator
upstream and a flanking $O1$ operator, -35 region and AseI site
downstream. This PCR product was combined with the appropriate
digest of pZS25'-YFP to give raise to
pZS25'$Oid$-E89-$O1_{-45}$-YFP. This is shown schematically in
fig.~\ref{fig:lacUV5O1-45}(B). Finally, the different lengths used
by Cloutier and Widom were generated from this template using site
directed mutagenesis.

\begin{figure}
\begin{center}
    \includegraphics[height=2.0truein]{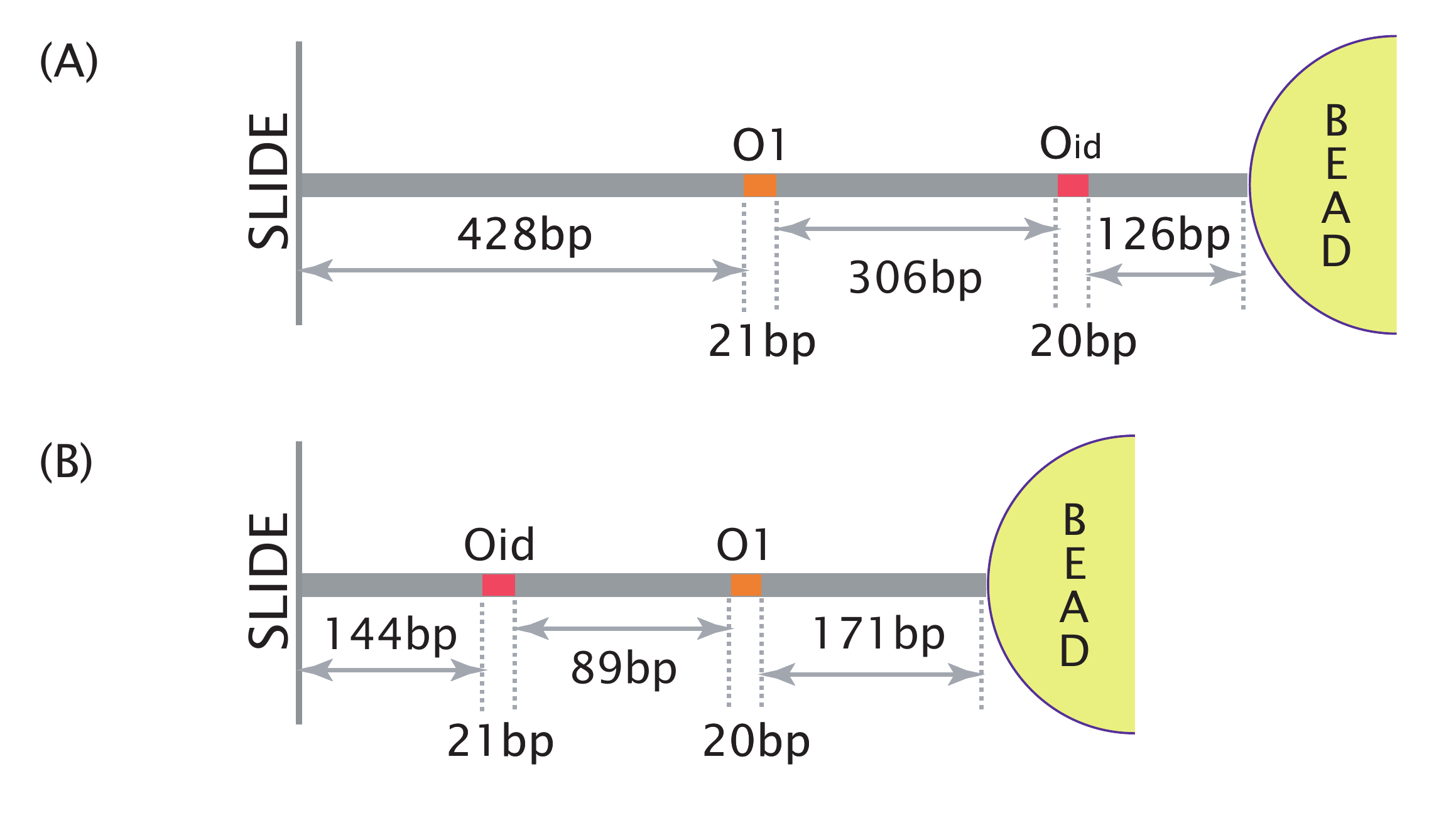}\\
    \caption{\small{Examples of the tether constructs used. (A) In the long distance constructs $Oid$ was displaced
    keeping the total construct length constant. (B) In the short distance constructs the sequence between the
    operators was altered, which results in each construct having a slightly different total length.
    (Drawings not to scale.)}}\label{fig:ConstructScheme}
\end{center}
\end{figure}

\subsection{ Construction of labeled DNAs}

In TPM experiments, DNA is linked between the substrate  and a bead.
Two pairs of linkers:  biotin-streptavidin and
digoxigenin-anti-digoxigenin, are chosen to permit specific linkage
of the DNA to a polystrene microsphere and glass coverslip,
respectively. As illustrated in fig.~\ref{fig:materials}(B), PCR was
used to amplify such labeled DNA with two modified primers. Each
primer is designed to be about 20~bp  in length and linked with
either biotin or digoxigenin at the 5' end (Eurofins MWG Operon). In
the case of the long sequence constructs, in order to optimize the
PCR reaction linearized plasmids with an AatII cut are used as the
template. Detailed information concerning the design of our PCR
reactions is listed in Table \ref{tab:pcrprimers} and the constructs
are shown schematically in fig.~\ref{fig:ConstructScheme}. The PCR
products were then purified by gel extraction (QIAquick Gel
Extraction Kit, QIAGEN) and the concentration of the DNA was
measured using quantitative DNA electrophoresis.

\begin{figure}
\begin{center}
\includegraphics[width=5in]{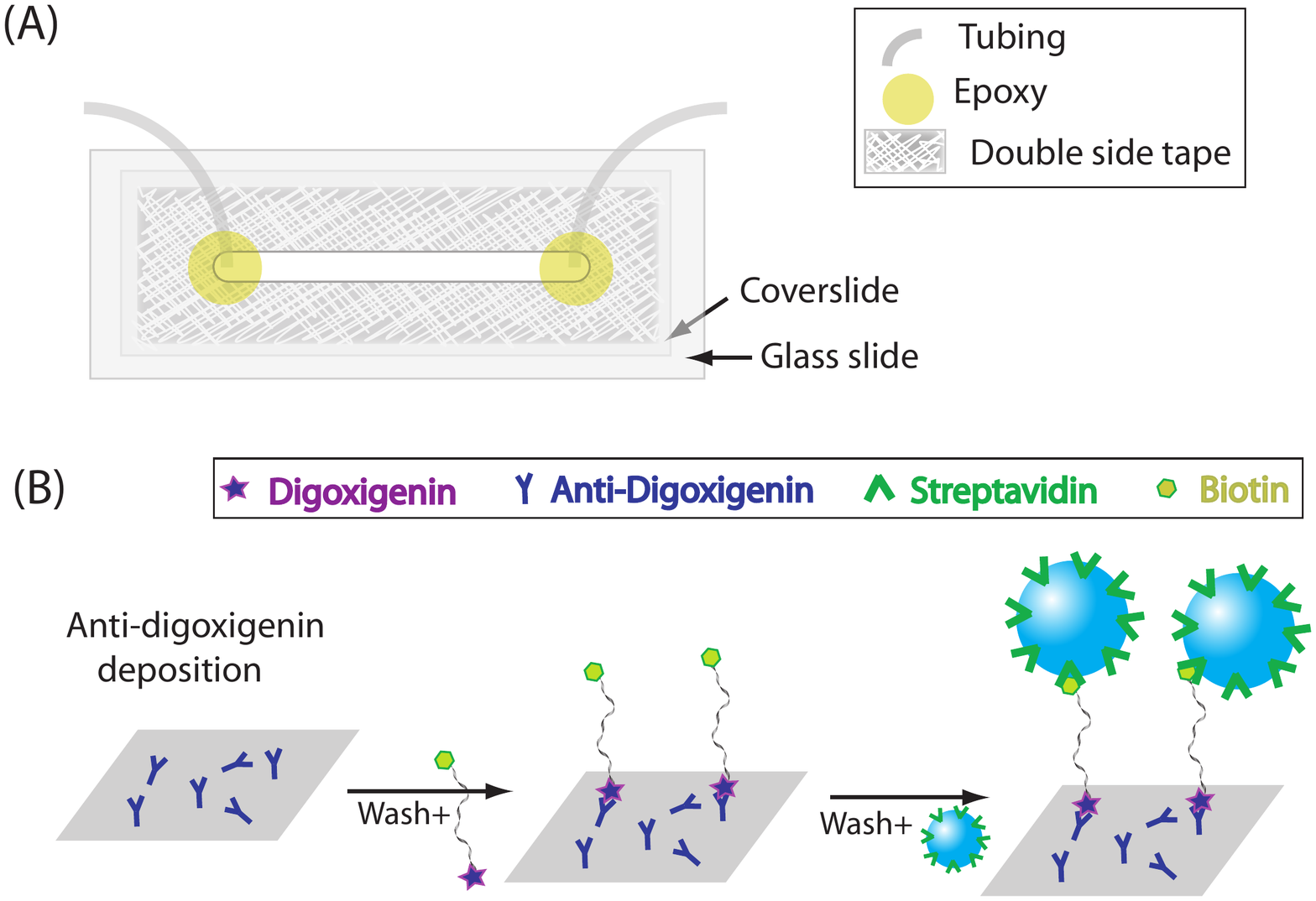}
\caption{\small{Illustration of TPM sample preparation.  (A) Sketch of the flow cell.  (B) The scheme for making DNA tethers. }}
\label{fig:tethers}
\end{center}
\end{figure}

\subsection{ TPM sample preparation}
TPM sample preparation involves assembly of  the relevant DNA
tethers and their associated reporter beads.  Streptavidin  coated
microspheres (Bangs lab) of diameter 490 nm served as our tethered
particle.  Prior to each usage, a buffer exchange on the beads was
performed by three cycles of centrifugation and resuspension in TPB
buffer (20~mM Tris-acetate, pH=8.0, 130~mM KCl, 4~mM MgCl$_{2}$,
0.1~mM DTT, 0.1~mM EDTA, 20 ~$\mu$g/ml acetylated BSA
(Sigma-Aldrich), 80 ~$\mu$g/ml heparin(Sigma-Aldrich) and 0.3$\%$
biotin-free casein colloidal buffer (RDI, Flanders, NJ)).  This
combination of reagents was chosen in an attempt to maximize sample
yield and longevity, while minimizing non-specific adsorption of DNA
and microspheres onto the coverslip.

Tethered particle samples were created inside a 20-40 $\mu$l flow
cell made out of a glass slide with one hole near each end, glass
coverslip, double-sided tape and tygon tubing.  The coverslip and
glass slide were cleaned with plasma cleaning for 4 minutes and then
the flow cell was constructed as shown in fig.~\ref{fig:tethers}(A).
Two tygon tubes serving as an input and output were inserted into
the holes on the glass slide and sealed with epoxy.  A reaction
chamber was created by cutting a channel on the double sided tape,
which glues the coverslide and glass slide together. Making the end
of the channel round and as close to the holes of the glass slide as
possible is important to avoid generating bubbles.  The flow cell
was then heated for about  20 seconds to seal securely.

For DNA tether assembly, the flow chamber was first incubated with
20 $\mu$g/mg polyclonal  anti-digoxigenin (Roche) in PBS buffer for
about 25 minutes, and then rinsed with 400 $\mu$l wash buffer (TPB
buffer with no casein)  followed by 400 $\mu$l of TPB buffer. 250
$\mu$l of labeled DNA in TPB buffer with about 2 pM concentration
was flushed into the chamber and incubated for around 1 hour.   Then
~10 pM of beads was introduced into the chamber and incubated for 20
minutes after washing with 750 $\mu$l TPB buffer to remove the
unbound DNAs.  Finally, unbound microspheres were removed by
flushing the chamber with 1 mL TPB buffer.  For looping experiments,
0.5 mL$\sim$ 1mL LRB buffer (10 mM Tris-Hcl, pH 7.4, 200~mM KCl, 0.1
~mM EDTA, 0.2 mM DTT, 5$\%$ DMSO and 0.3$\%$ biotin-free casein
colloidal buffer (RDI, Flanders, NJ)) containing the desired
concentration of Lac repressor  (a kind gift from Kathleen Matthews'
lab) was then flushed into the chamber and incubated about 15
minutes before observation.

\begin{table*}
\begin{center}
\begin{tabular}{|c|ccc|c|}
 \hline  Molecule        &Primer         &Template                 &Action                   & Resulting Molecule        \\  \hline
    pUC19O1         & Mut0                           &pUC19         &Replace         &O1   \\
    pUC300         & Mut1                           &pUC301         &Delete 1bp      &O1-300bp-Oid   \\
    pUC301         & Mut2                           &pUC19O1         &Replace       &O1-301bp-Oid   \\
    pUC302         & Mut3                           &pUC19O1         &Replace       &O1-302bp-Oid   \\
    pUC303         & Mut4                           &pUC19O1         &Replace       &O1-303bp-Oid   \\
    pUC304         & Mut5                           &pUC19O1         &Replace       &O1-304bp-Oid   \\
    pUC305         & Mut6                           &pUC19O1         &Replace       &O1-305bp-Oid   \\
    pUC306         & Mut7                           &pUC19O1         &Replace       &O1-306bp-Oid   \\
    pUC307         & Mut8                           &pUC19O1         &Replace       &O1-307bp-Oid   \\
    pUC308         & Mut9                           &pUC19O1         &Replace       &O1-308bp-Oid   \\
    pUC309         & Mut10                         &pUC308           &Add 1bp               &O1-309bp-Oid   \\
    pUC310         & Mut11                         &pUC308           &Add  2bp             &O1-310bp-Oid   \\  \hline
\end{tabular}
\end{center}

 {\small{Primer sequences(5' -$>$ 3'):\\
 Mut0: ctaactcacattaattgcgttgAgctcGAGgTTcgctttccagtc\\
 Mut1: catacgagccggaa (G) cataaagtgtaaagc\\
 Mut2: ctcggaaagaaca AATTGTGAGCGCTCACAATT aaggccaggaacc   \\
 Mut3: ctcggaaagaacat AATTGTGAGCGCTCACAATT aggccaggaaccg   \\
 Mut4: cggaaagaacatg AATTGTGAGCGCTCACAATT ggccaggaaccgt   \\
 Mut5: ggaaagaacatgt AATTGTGAGCGCTCACAATT gccaggaaccgta   \\
 Mut6: gaaagaacatgtg AATTGTGAGCGCTCACAATT ccaggaaccgtaa    \\
   Mut7:  cggaaagaacatgtga AATTGTGAGCGCTCACAATT caggaaccgtaaaaag    \\
   Mut8:  ggaaagaacatgtgag AATTGTGAGCGCTCACAATT aggaaccgtaaaaagg   \\
   Mut9:  gaaagaacatgtgagc AATTGTGAGCGCTCACAATT ggaaccgtaaaaaggc   \\
   Mut10: catacgagccggaag [C]  cataaagtgtaaagc   \\
   Mut11: catacgagccggaag [CG] cataaagtgtaaagc }}  \\

\caption{\small{ Materials used in the mutagenesis process for
creating plasmids with two Lac repressor binding sites. The capital
letters in the primer sequences indicate the mutations. '()'
indicates bp deletion and '[ ]' indicates bp addition. The
inter-operator distance indicated here is the distance between two
inner edges of the operators instead of center to center distance
that is commonly used in {\it in vivo} experiments
\citep{Oehler1990,Oehler1994,Muller1996,Becker2005,Becker2007}.}}
\label{tab:mutprimers}
\end{table*}

\begin{table*}
\begin{center}
\begin{tabular}{|cccc|}
 \hline  Molecule        &Template      &Length(bp)                   & Resulting        \\  \hline
    pUC300L1              &pUC300         &900       &Dig - 427bp-O1-300bp-Oid-132bp - Bio   \\
    pUC301L1              &pUC301         &901       &Dig - 427bp-O1-301bp-Oid-132bp - Bio   \\
    pUC302L1              &pUC302         &901       &Dig - 427bp-O1-302bp-Oid-131bp - Bio    \\
    pUC303L1              &pUC303         &901       &Dig - 427bp-O1-303bp-Oid-130bp - Bio   \\
    pUC304L1              &pUC304         &901       &Dig - 427bp-O1-304bp-Oid-129bp - Bio   \\
    pUC305L1              &pUC305         &901       &Dig - 427bp-O1-305bp-Oid-128bp - Bio   \\
    pUC306L1              &pUC306         &901       &Dig - 427bp-O1-306bp-Oid-127bp - Bio    \\
    pUC307L1              &pUC307         &901       &Dig - 427bp-O1-307bp-Oid-126bp - Bio   \\
    pUC308L1              &pUC308         &901       &Dig - 427bp-O1-308bp-Oid-125bp - Bio    \\
    pUC309L1              &pUC309         &902       &Dig - 427bp-O1-309bp-Oid-125bp - Bio    \\
    pUC310L1              &pUC310         &903       &Dig - 427bp-O1-310bp-Oid-125bp - Bio    \\  \hline
    E8-89                 &pZS25'$Oid$-E89-$O1_{-45}$-YFP   &445       &Dig - 144bp-Oid-89bp-O1-171bp - Bio   \\
    E8-94                 &pZS25'$Oid$-E94-$O1_{-45}$-YFP   &450       &Dig - 144bp-Oid-94bp-O1-171bp - Bio    \\
    E8-100                &pZS25'$Oid$-E100-$O1_{-45}$-YFP  &456       &Dig - 144bp-Oid-100bp-O1-171bp - Bio   \\   \hline

\end{tabular}
\end{center}
{\small{ Primer sequences(5' -$>$ 3'):\\
 Plen901F:  Dig - ACAGCTTGTCTGTAAGCGGATG \\
 Plen901R: Bio - CGCCTGGTATCTTTATAGTCCTGTC}   \\
 PF1: Dig - ATGCGAAACGATCCTCATCC    \\
 PR1: Bio - GCATCACCTTCACCCTCTCC
\caption{\small Materials used in amplifying labeled DNA using PCR.
The inter-operator distances indicated here is the distance between
two inner sides of the operators instead of center to center
distance. Primers Plen901F and Plen901R were used for the long
distance constructs. Primers PF1 and PR1 were used for the short
distance constructs.}} \label{tab:pcrprimers}
\end{table*}

\subsection{Data Acquisition and Processing}
The motion of the bead is recorded through a Differential
Interference Contrast (DIC) microscope at 30 frames per second.  The
position of the bead is tracked in the x-y plane using a
cross-correlation method \citep{Gelles1988} and recorded as raw data
for further analysis.  Such raw positional data are subject to  a
slow drift due to vibrations of the experimental apparatus.  A drift
correction is then applied using a  high pass first-order
Butterworth filter at cutoff frequency 0.1Hz~\citep{Vanzi2006}. From
the filtered data, $R^2(t)$ is then calculated as $x(t)^2+y(t)^2$
and a running average $\sqrt{<R^2(t)>}$ is obtained using a Gaussian
filter at cutoff frequency 0.033 Hz~\citep{Vanzi2006,
Colquhoun1995}, which corresponds to the standard deviations of the
filter's impulse response time of 4~s. The traces shown in this
paper are all obtained in this way.

\noindent{\bf Acknowledgements}  We are extremely grateful to a
number of people who have been generous with their ideas, time and
materials.  Jon Widom has given us help of all kinds since we first
began down the path of trying to make these measurements.   Jeff
Gelles has continuously helped us along with insights from the very
beginning when one of us (RP) didn't even know what DIC microscopy
was.  Similarly, Laura Finzi and David Dunlap were kind enough to
host one of us (LH) in their lab and to provide constant
encouragement, advice and insights.     Bob Schleif has patiently
advised us on many aspects of this project and has served as a
looping sage for many years.   Kathleen Matthews and Jason Kahn have
both been extremely generous with both ideas and in providing us
protein. We also thank Sankar Adhya, David Bensimon, Nily Dan, Paul
Grayson, Heun Jin Lee, John Maddocks, Keir Neuman, Tom Perkins,
Steve Quake, Andy Spakowitz, Terence Strick, Paul Wiggins, Jie Yan,
and Sylvain Zorman for many discussions. RP and LH acknowledge the
support of the Keck Foundation, National Science Foundation grant
Nos. CMS-0301657 and CMS-0404031, and the National Institutes of
Health Director's Pioneer Award grant No. DP1 OD000217. HG is
grateful for support from both the NSF funded NIRT and the NIH
Director's Pioneer Award. PCN, KBT, and JFB were partially supported
by NSF grants DGE- 0221664, DMR04-25780, and DMR-0404674. \clearpage

\section{Supplementary Materials}

\subsection{Bead Selection, Data Rejection and ``Representative Data''}

\begin{figure}
\begin{center}
\includegraphics[width=4.0in]{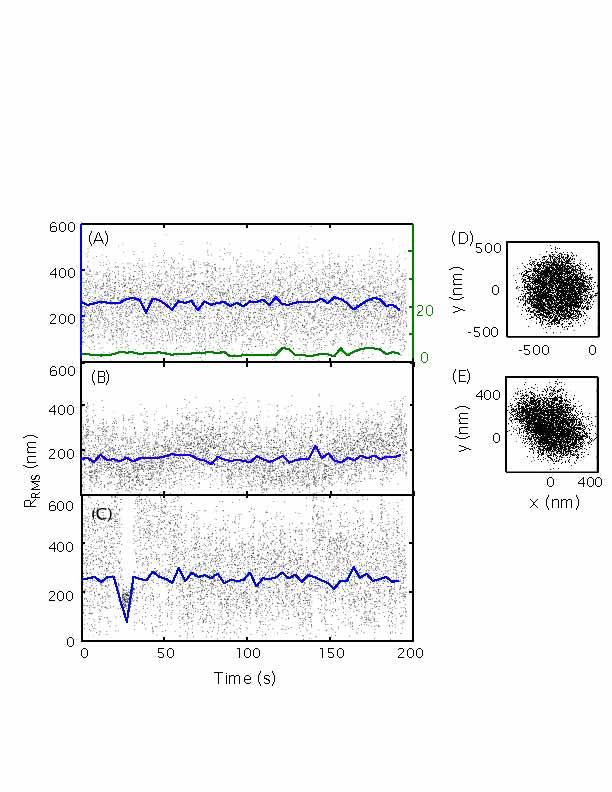}
\caption{\small{Conceptual description of data selection.   All traces in this case are taken
in the absence of Lac repressor and are used as the basis of
choosing qualified beads for the looping study.   (A) Experimental traces for a
bead exercising symmetric motion (blue) and for a stuck bead (green).  (B)  Trajectory for
a bead that exhibits non symmetric motion. (C) Trajectory for a bead that exhibits a
transient sticking event.  (D) Positional data for a bead that exhibits symmetric
motion.  (E) Positional data corresponding to the trajectory shown in (B) and for which
the motion is not symmetric.}}
\label{DataSelection}
\end{center}
\end{figure}

\begin{figure}
\begin{center}
\includegraphics[width=4.0in]{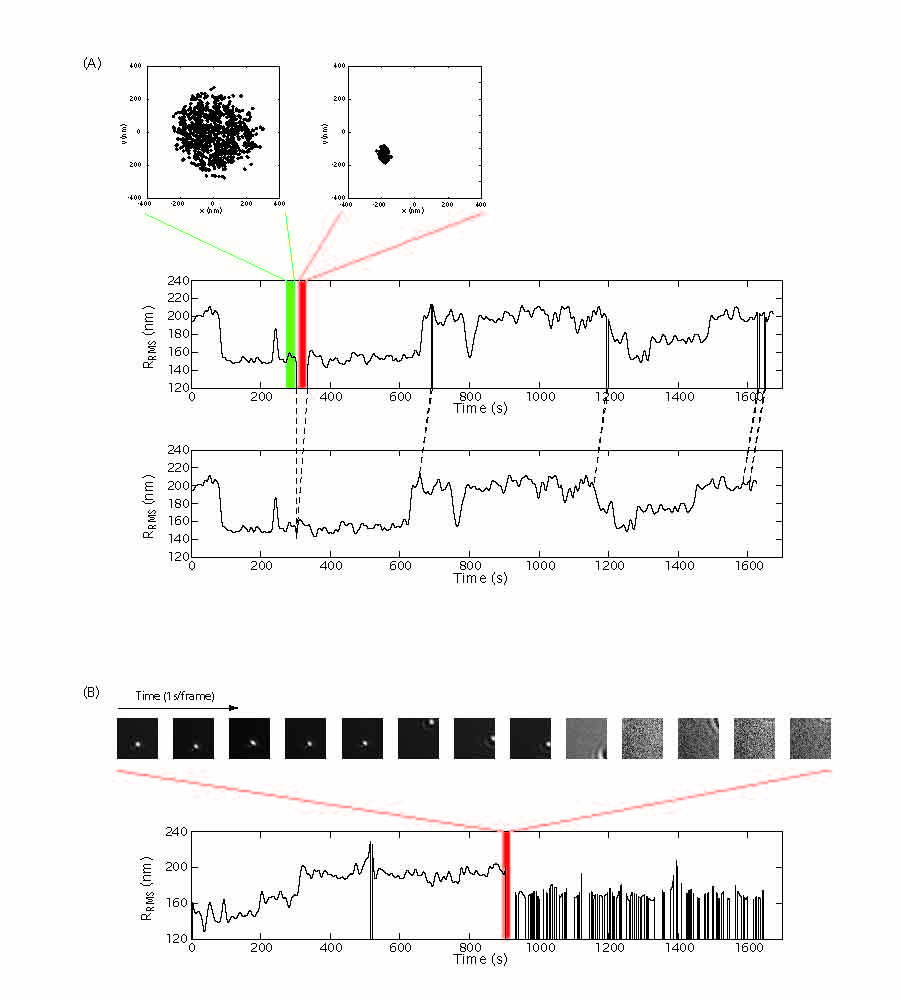}
\caption{\small{Transient sticking events and tether breaking. (A) A transient
sticking event is revealed by a dramatic reduction in the movement
of the bead and is associated with a spike in the R$_{\mbox{RMS}}$
trace. These ``offending'' regions of the traces can be excised out which will not
affect the resulting histogram, but might present an issue for any kinetic analysis as discussed
in the text. A movie corresponding to this event is provided as a Supplementary Movie.
(B) Signature of a tether breaking.}} \label{RejectedData}
\end{center}
\end{figure}

One of the most important challenges of these experiments (and
perhaps any single-molecule experiment based upon watching the
motions of beads tethered to single molecules)  is devising
systematic methods for deciding which beads are ``qualified'' and
how to reject trajectories that are anomalous without biasing the
results \citep{Pouget2004,Blumberg2005,Pouget2006,Nelson2006}. To
that end, we have attempted to institute a number of criteria for
performing data selection that are indicated schematically in
figs.~\ref{DataSelection} and \ref{RejectedData}. The first attempt
to ``objectively'' select qualified beads takes place by excising
segments of the traces corresponding to the unlooped state and
examining whether their motions are symmetric (i.e. jiggle in the x-
and y- directions in the same way) as evidenced by the probability
distribution for the x- and y- excursions.  This screening permits
us to select beads within a given field of view that are ostensibly
properly tethered. Examples of these selection criteria are shown in
fig.~\ref{DataSelection} for the particular case where no protein is
present. Typically, a fraction of roughly $20~\sim~30~\%$ of the
beads are rejected as a result of failure to exhibit proper symmetry
or because they are stuck.

A more tricky question arises when we have to assess whether
something went wrong during data acquisition that requires either
all or part of a given TPM trajectory to be rejected.  In some
cases, the offending behavior is evident at the level of the bare
images of the jiggling beads.  For example, a given bead can become
stuck to the surface or the DNA can break and the bead will
disappear from the field of view. These events have a signature of
spikes in the R$_{\mbox{RMS}}$ traces as shown in
fig.~\ref{RejectedData}. A movie corresponding to the event shown in
fig.~\ref{RejectedData}(A) can be found as a Supplementary Movie.

Fig.~\ref{RejectedData} also shows an example of data that was kept
with an offending region highlighted that was removed. Note that if
the spike regions in trajectories were actually kept, it would have
no bearing on histograms like those shown in
figs.~\ref{fig:histogram} and \ref{fig:histogrambp} since the spikes
will show up as features on the tails of the histograms.  On the
other hand, by excising certain pieces of trajectories, there can be
some effect on the kinetic claims we would be able to make since
these anomalies will cause errors in the dwell time measurements.

In none of the cases considered in this work were sticking events
observed in any significant number. Assuming that sticking is mainly
due to nonspecific interactions with the bead and the surface one
would expect the shorter constructs to show the most sticking
events. In order to control for this we performed TPM experiments
using tethers of 351~bp in length in the absence of Lac repressor.
This length is comparable to the length the short constructs (E889,
E894 and E8100) would have if the sequence between the {\it lac}
operators was removed. Out of 18 tethers characterized only 5 showed
any sticking events. In those 5 traces, the sticking events
corresponded to less than 4~\% of the observation time for each bead
(data not shown). In order to discard any contribution to the
sticking events from the presence of the protein, Lac repressor was
flowed in in the presence of 1~mM IPTG which serves to eliminate the
binding of Lac repressor to the DNA (or at least drastically reduce
it).  The goal of this control is to see whether the presence of
unbound protein somehow induces unwanted sticking events. Out of the
7 tethers characterized all showed sticking events, but these
corresponded to less than 1~\% of the time. Finally, there is still
the chance that Lac repressor that is specifically bound to the
tether might contribute to sticking. In order to test this
hypothesis we used a construct of this length with only one binding
site.   Here too (data not shown), there was no significant sticking
lending further support for the idea that even for the short
tethers, we are able to detect looping.

\begin{figure}
\begin{center}
\includegraphics[width=5in]{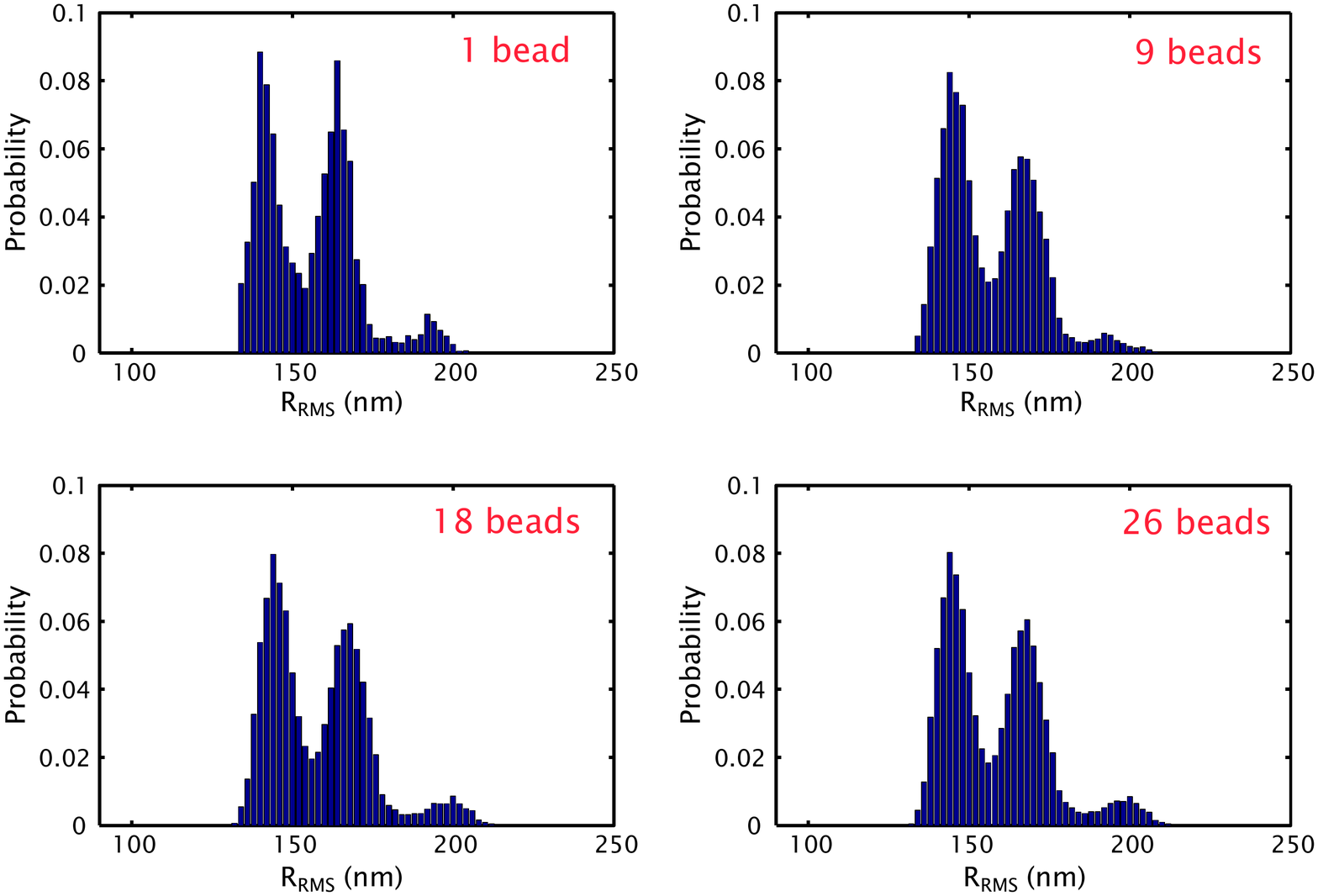}
\caption{\small{Effect of averaging on the data.   These four histograms show the effects of including different numbers
of beads in determining the overall average.   Data obtained with pUC306L1 DNA in the presence of 10~pM Lac repressor.}}
\label{fig:avgbar}
\end{center}
\end{figure}

In order to produce histograms like those shown in
figs.~\ref{fig:histogram} and \ref{fig:histogrambp} we have to sum
over the histograms resulting from many individual trajectories.
Fig.~\ref{fig:trajectory} shows the connection between an individual
TPM trace for a single bead and its corresponding motion histogram.
However, since each trajectory has its own unique features, it is of
interest to see how the smoothed histogram resulting from many
individual trajectories emerges from the averaging process.
Fig.~\ref{fig:avgbar} shows the motion histogram obtained by
averaging over the histograms from progressively larger numbers of
individual trajectories.

\begin{figure}
\begin{center}
\includegraphics[width=5in]{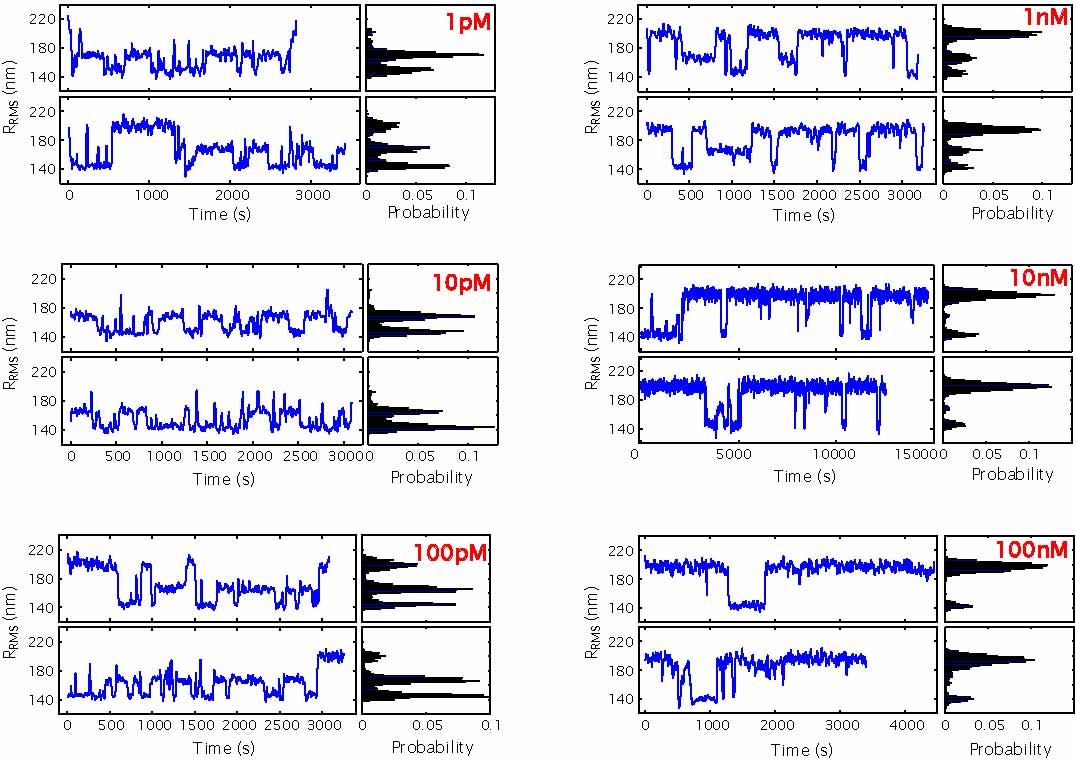}
\caption{\small{Concentration dependence of TPM trajectories. Representative examples of TPM
trajectories.  Typical TPM trajectories of the DNA tethered beads in
the presence of different concentrations of Lac repressor varying
from 1~pM to 100~nM. The total DNA length is 901~bp and the interoperator spacing is 306~bp.}}
\label{fig:alltraces}
\end{center}
\end{figure}

\begin{figure}
\begin{center}
\includegraphics[width=4in]{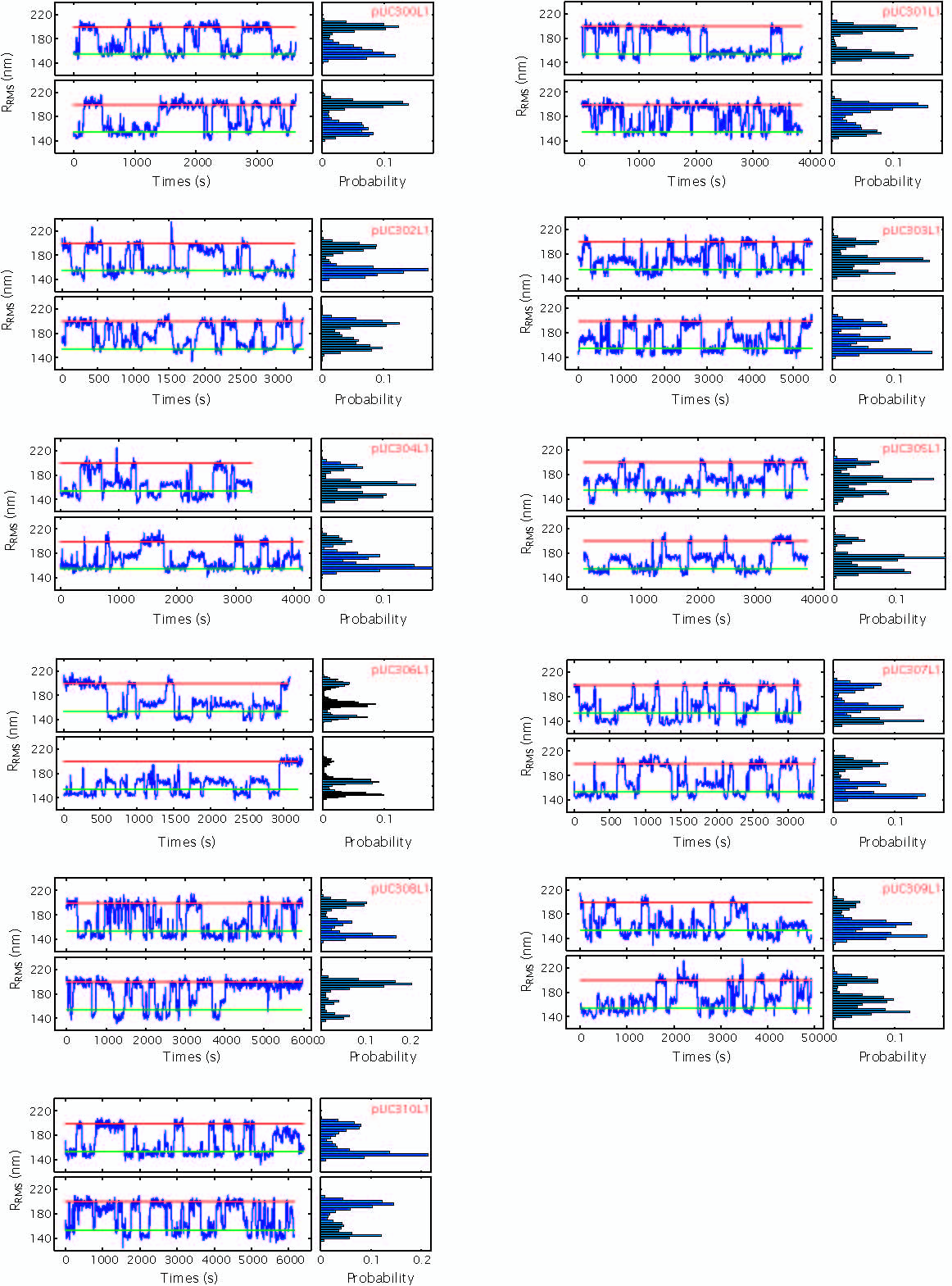}
\caption{\small{Length dependence of TPM trajectories.  Typical TPM trajectories
 of the DNA tethered beads with interoperator spacing from 300 to 310 bp in 1~bp increments.  The
 concentration of Lac repressor used in this set of experiments  was 100~pM.  The distance between the
 two operators is indicated in the naming of the construct.}}
\label{fig:lengthalltraces}
\end{center}
\end{figure}

\begin{figure}
\begin{center}
\includegraphics[width=3in]{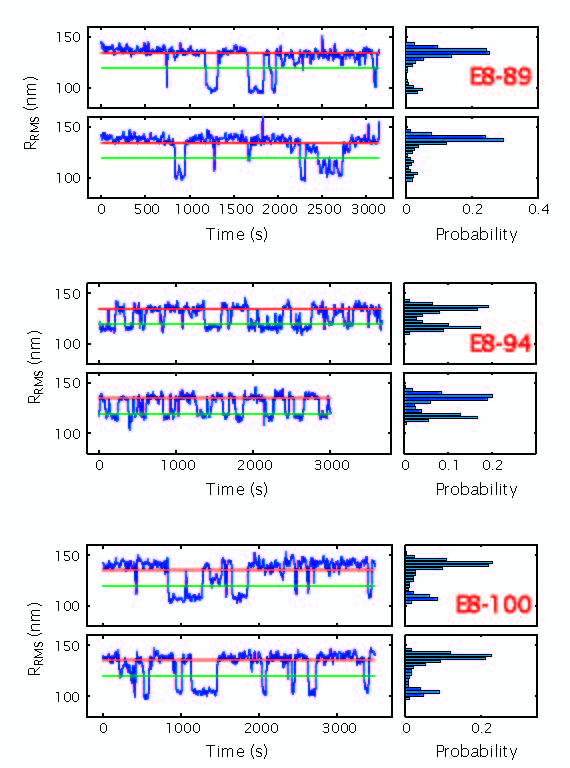}
\caption{\small{Typical TPM trajectories for DNA tethered beads with interoperator spacing of
89~bp, 94~bp and 100~bp.  E8 refers
to the particular sequence used in these experiments.   The concentration of  Lac repressor used
to generate these trajectories is 100pM.  The red
and green lines indicate the expected excursion for the unlooped and looped states, respectively, where
the expected length of the looped state is based upon subtracting the interoperator spacing from the
overall tether length.}}
\label{fig:E8trace}
\end{center}
\end{figure}

Now that we have seen some of the pitfalls associated with TPM
trajectories, we show ``representative'' examples of the individual
trajectories culminating in  figs.~\ref{fig:histogram} and
\ref{fig:histogrambp}.  Fig.~\ref{fig:alltraces} shows multiple
examples of trajectories resulting from different concentrations of
Lac repressor.  Even at the level of visual inspection of these
individual trajectories, it is evident that there are two distinct
looping states and that the relative occupancies of the different
looped and unlooped states depend upon the concentration of
repressor.  Similar results are shown in
figs.~\ref{fig:lengthalltraces} and \ref{fig:E8trace} which
illustrates multiple individual trajectories for the case in which
the interoperator spacing (rather than the Lac repressor)
concentration is the experimental dial that we tune to vary the
looping stability.

\subsection{Data Analysis and Probabilities
Calculation}\label{sec:Analysis}

The data shown in  figs.~\ref{fig:histogram} and
\ref{fig:histogrambp} characterizes the results of many different
TPM trajectories for each condition (Lac repressor concentration or
interoperator spacing). We are interested in obtaining the
probabilities associated with each state and to that end, we have
tried a variety of different approaches to examine the sensitivity
of the results to method of data analysis.

The first analysis we explored is based on directly looking at
histograms such as those shown in figs.~\ref{fig:histogram} and
\ref{fig:histogrambp}. As mentioned in the main text, these
histograms are the result of adding up the normalized contribution
from each bead. One scheme for carrying this out is to fit the
histogram to the sum of three Gaussians. The idea of such a fit is
that there is a main peak associated with the unlooped state and
then two separate looping peaks, each of which is fit with its own
Gaussian. With the fitting results in hand the area under each
Gaussian can be computed, which leads to a probability assignment.
We call this scheme ``Gaussian Integral''.

An alternative scheme is to define thresholds between the different
states. The bins on either side of the thresholds are then added,
giving the different probabilities. We explore two ways of
calculating the thresholds: i) Finding the minimum between adjacent
Gaussians from the fit described previously (``Gaussian Minimum''),
and ii) finding the minimum in the histograms between peaks
(``Histogram Minimum'').

Finally, we have also explored the use of alternatives such as the
Diffusive Hidden Markov Model (``dHMM''). The Diffusive Hidden
Markov Model (DHMM) method is applied to do the kinetic
analysis~\citep{Beausang2007,beau2007} and for our present purposes
permits us a different way to determine the looping probability by
telling us the fraction of time spent in each of the distinct
states.  This method employs the concept of HMM and customizes it in
a way suitable  for TPM data, through which the rate constants are
directly derived from the positional data obtained in the TPM
experiments.  To characterize the dynamical information of the beads
in each state,  control experiments are performed in the following
ways:  i) To obtain the information for the unlooped state,  the
bead's motion is observed in the absence of the DNA looping protein
Lac repressor.  ii) For the looped state, we monitor the bead's
motion in the presence of a Lac repressor mutant V52C instead of Lac
repressor itself.  This mutant is designed to permit disulfide bond
formation, which makes important contacts that are critical to DNA
binding. As a result, V52C has increased affinity for DNA
operators~\citep{Falcon1997}, leading to a measurement of primarily
looped states. Such data containing only one type of looped state is
selected to obtain the information that serves as input to the HMM
model.  One of the outcomes of the HMM analysis is an explicit
statement about the amount of time spent in each of the states which
can be used in turn to compute the looping probability.

\begin{figure}
\begin{center}
    \includegraphics[height=4.0truein]{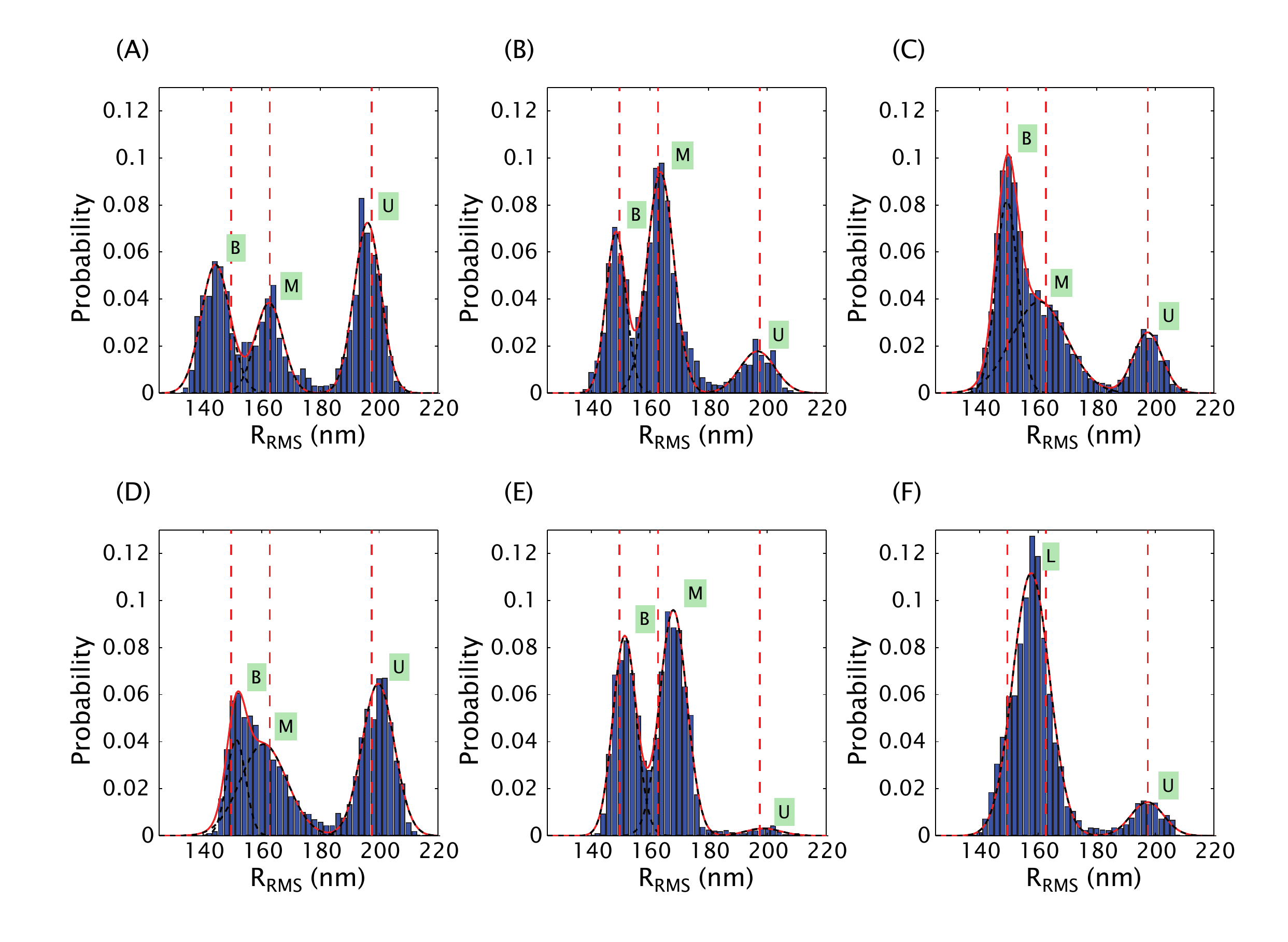}\\
    \caption{\small{Rogues gallery of individual bead histograms. Three Gaussian fit to individual
    bead traces corresponding to the pUC300L1 construct. The vertical dashed lines correspond to
    the locations of the peaks as revealed by a three Gaussian fit to the corresponding histogram
    of fig.~\ref{fig:histogrambp}. The black dashed line are the individual Gaussians, while the
    solid red line is their sum. (A-E) The peaks are labeled B (bottom loop), M (middle loop), and
    U (unlooped state). In the small fraction of cases that no decision about the identity of
    the looped state could not be discerned the label L (looped state) is used.}}\label{fig:300bpGallery}
\end{center}
\end{figure}

One argument against the previously mentioned schemes is that they
do not capture the variability inherent in single molecule
experiments. Each tether will behave in a slightly different way, as
is illustrated in fig.~\ref{fig:300bpGallery} for construct
pUC300L1. Notice that even though the two looped states were
overlapping in fig.~\ref{fig:histogrambp} they are discernable in
most individual traces. Fig.~\ref{fig:300bpGallery}(F) also shows a
case where no call on the identity of the looped state could be
made. For the long length constructs where this happened only a
small fraction of the beads, between 2\% and 6\% would show this
type of histogram. Identification of the individual loops becomes
more problematic in the short length constructs. In this case around
10\% of the beads would show this behavior.

\begin{figure}
\begin{center}
    \includegraphics[height=2.0truein]{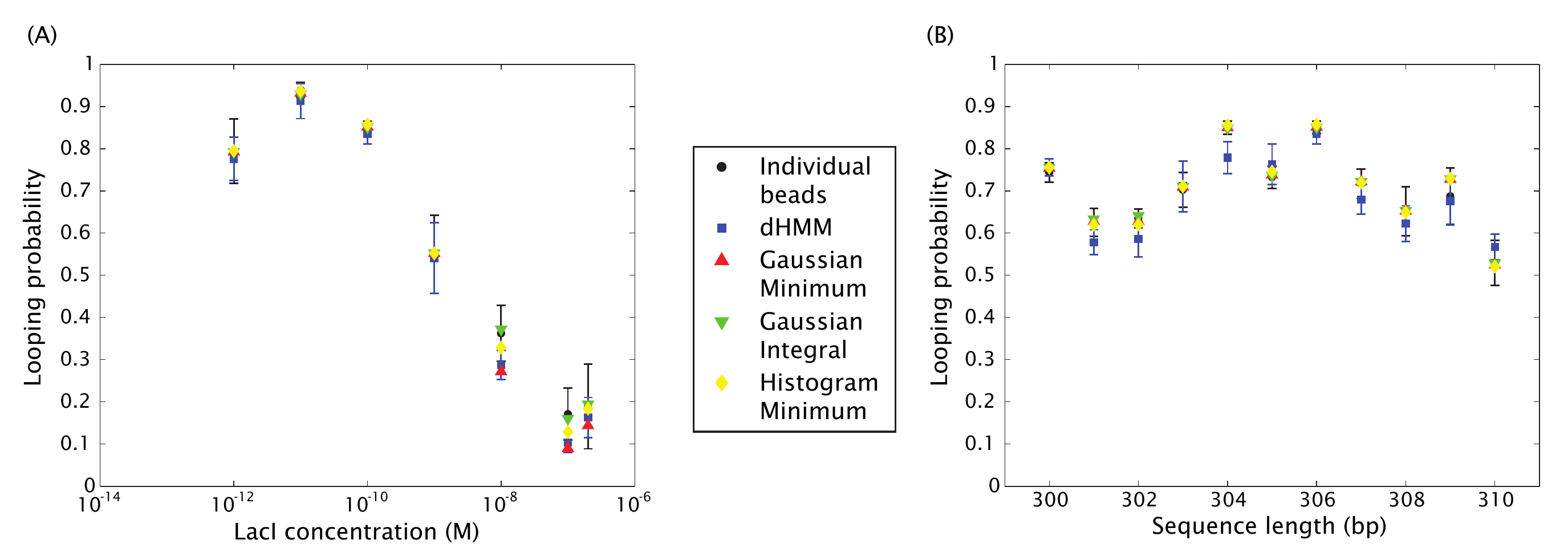}\\
    \caption{\small{Different approaches for calculating the looping probability. The looping probability
    as a function of  (A) concentration and  (B) sequence length,  calculated using the approaches described in
    the text.}}\label{fig:pLoopComparison}
\end{center}
\end{figure}

\begin{figure}
\begin{center}
    \includegraphics[height=2.0truein]{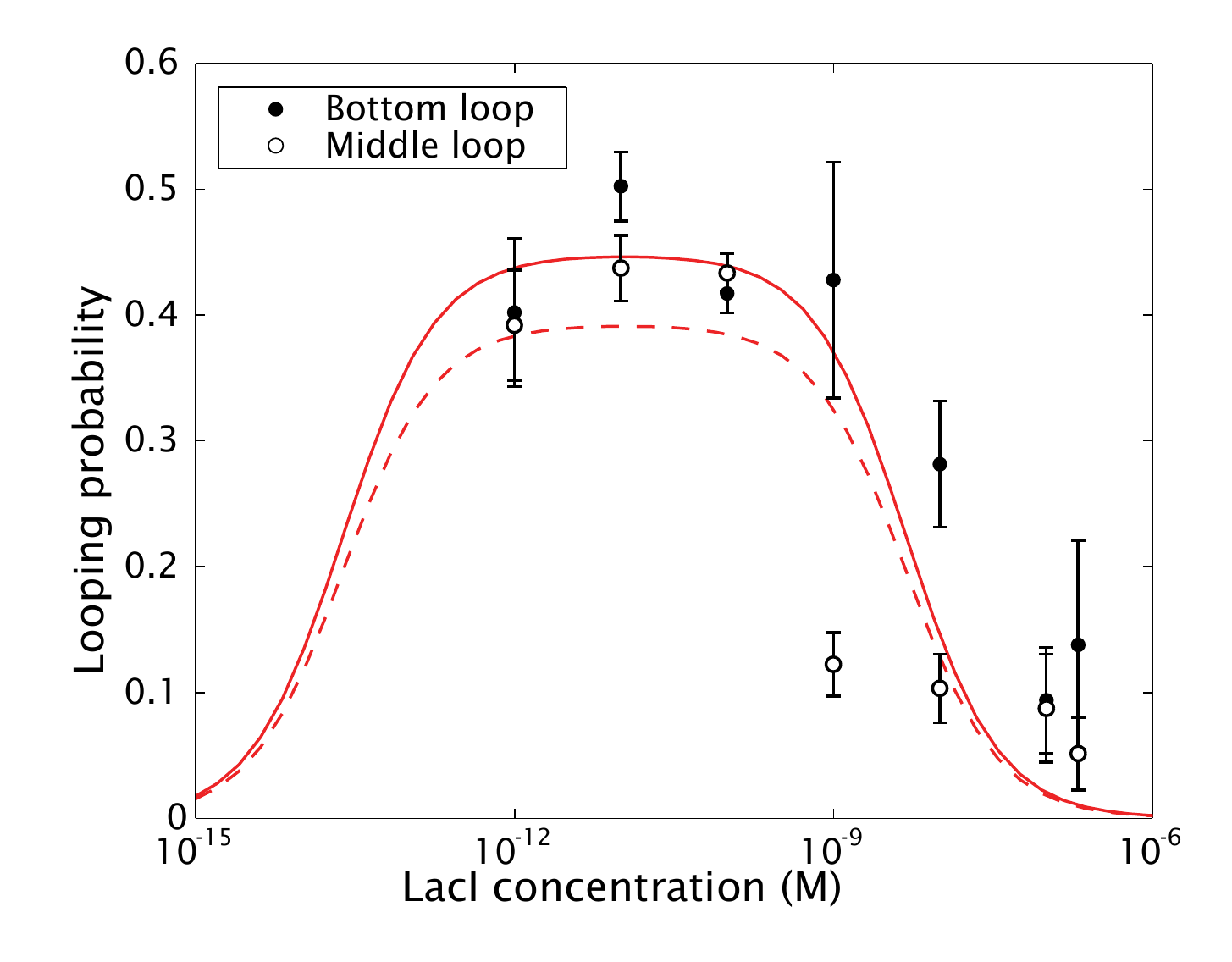}\\
    \caption{\small{Individual loops vs. concentration. Probability of each looped state as a function
    of concentration. The lines are fits to the nonlinear model from eqn.~\ref{eq:ploopThermoIndiv}.}}\label{fig:pLoopConcIndividual}
\end{center}
\end{figure}

\begin{figure}
\begin{center}
    \includegraphics[height=2.0truein]{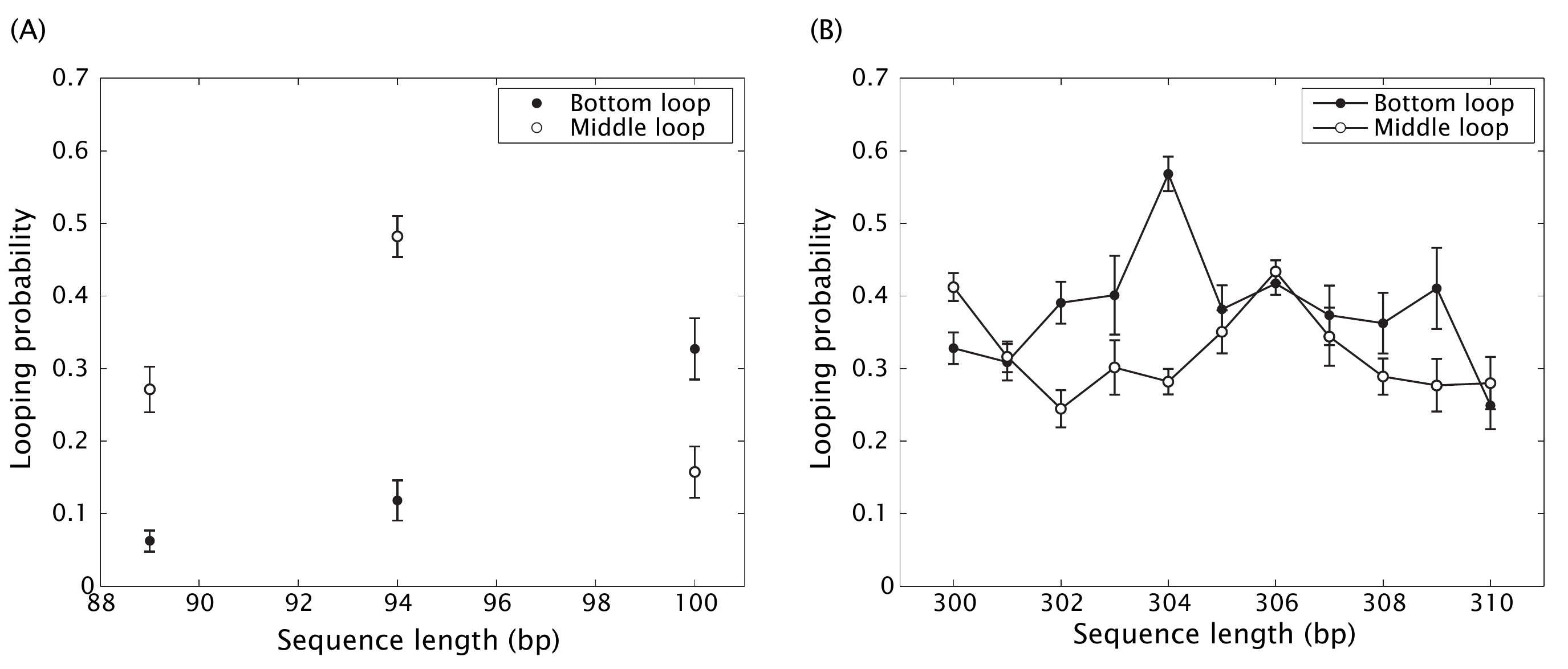}\\
    \caption{\small{Individual loops vs. phasing. Probability of each looped state as a function
    of sequence length. (A) Short loops, (B) a full cycle at 300~bp.}}\label{fig:pLoopLengthIndividual}
\end{center}
\end{figure}

The looping probabilities obtained using all these methods are shown
in fig.~\ref{fig:pLoopComparison}. We conclude that there is no
significant variation in the results from any of the different
approaches. In section \ref{sec:TheoryVsExperiment} we show that the
quantitative parameters extracted from these different looping
probabilities do not differ significantly. Finally,
figs.~\ref{fig:pLoopConcIndividual} and
\ref{fig:pLoopLengthIndividual} show the looping probability for
each individual state in the cases where both states were
discernible.  Ultimately, it would be of great interest to use
experiments like those described here to determine the looping free
energies (or $J_{loop}$s for the different states. This is presented
in section \ref{sec:IndivLoops}.

\subsection{Theoretical Analysis of Looping}

Statistical mechanics provides a powerful tool for dissecting the
DNA-protein interactions that take place during transcriptional
regulation.    We find it convenient to derive the various
expressions for binding probabilities using simple lattice models of
DNA binding proteins and their DNA targets.  These models can then
be reinterpreted in the familiar language of equilibrium constants
and effective $J$-factors.    In this section, we sketch the
derivations of the formulae used in the main body of the paper. An
alternative derivation appears in \citep{Towles2008}.

\subsubsection{Simple binding of Lac repressor}\label{sec:LacISimple}

In a lattice model, we imagine the solution as discretized into a
set of $\Omega$ boxes of volume $v$.  The $R$ repressors are free to
occupy any of these distinct boxes which provide a simple and
convenient basis for computing the entropic contribution to the
overall free energy. A repressor in solution has an energy
$\varepsilon_{sol}$ which appears in the Boltzmann factor. The
configurational degrees of freedom (both translational and
rotational) in this model are taken care of by assigning the
molecules to one of the $\Omega$ boxes available in our lattice
model of the solution and by noting that there is  a  factor of ${8
\pi^2 \over \delta \omega}$ associated with its rotational degrees
of freedom ($4 \pi$ for the directions in which the molecule can
point on the unit sphere  and $2 \pi$ for the rotation around the
protein's axis). The partition function of $R$ repressors in the
solution is
\begin{equation}
    Z_{sol} = {\Omega \choose R} e^{-\beta R \varepsilon_{sol}} \left({8 \pi^2 \over \delta \omega} \right)^R.
\end{equation}

Now we introduce a DNA molecule with one binding site. This case is
appropriate when LacI is in excess of the DNA. When one Lac
repressor from the solution binds to the operator it now has an
energy $\varepsilon_{b}$ associated with the binding itself and a
``tether'' energy $\varepsilon_{t}$ associated with the extra head
that is still in the solution. Next, we exploit the fact that we can
choose either head to bind to the operator of interest and this head
can bind in two distinct orientations, yielding a factor of 4
degeneracy in this state. The total partition function is
\begin{equation}
    Z = Z_{sol}(R) + 4 Z_{sol} (R-1) e^{-\beta (\varepsilon_{b}+\varepsilon_t)}.
\end{equation}
This translates into the following probability of binding
\begin{equation}
    p_{bound} = \frac{4\frac{\delta \omega}{8 \pi^2} \frac{R}{\Omega} e^{-\beta\Delta \varepsilon}}
                    {1+4\frac{\delta \omega}{8 \pi^2} \frac{R}{\Omega} e^{-\beta\Delta \varepsilon}},
\end{equation}
where we have defined $\Delta \varepsilon = \varepsilon_b +
\varepsilon_t - \varepsilon_{sol}$.

We recover  the usual formula when characterizing binding using
dissociation constants
\begin{equation}
    p_{bound} = \frac{[R]/K_d}{1+ [R]/K_d},
\end{equation}
if we make the identification
\begin{equation}\label{eq:Kd2Epsilon}
    K_{d} = \frac{1}{4v} \frac{8 \pi^2}{\delta \omega} e^{\beta \Delta\varepsilon}.
\end{equation}
With this result in hand we are ready to address the more complex
case of DNA looping.

\subsubsection{DNA looping by Lac repressor}

We now have two operators, each one with a binding energy
$\varepsilon_1$ and $\varepsilon_{id}$, corresponding to the
operators $O1$ and $Oid$, respectively. We consider the usual five
classes of states that include: i) free operators, ii+iii) one of
the operators occupied, iv) both operators occupied by different
LacI molecules, and v) LacI looping both operators, which can happen
in multiple configurations. The partition function is
\begin{eqnarray}
    Z &=& Z_{sol}(R) + 4 Z_{sol} (R-1)  e^{-\beta \varepsilon_t}
            \left( e^{-\beta \varepsilon_1} + e^{-\beta \varepsilon_{id}} \right)     \\ \nonumber
        &&+ 16 Z_{sol}(R-2)  e^{-\beta(\varepsilon_1 + \varepsilon_{id} + 2 \varepsilon_t)} + \\ \nonumber
        &&+ \sum_i Z_{sol}(R-1) e^{-\beta(\varepsilon_1 + \varepsilon_{id} + F_{loop,i})}. \nonumber
\end{eqnarray}
The factors of 4 in the second and third term correspond to the
degeneracy described above.  The factor of 16 in the fourth term
accounts for all of the different ways of binding two repressors
independently. Here we defined $F_{loop,i}$ as the looping free
energy associated with a particular configuration (orientation of
operators with respect to the molecule). The sum in the last term
includes all four possible loop topologies
\citep{Semsey2004,Swigon2006} and the fact that we are thinking of
the two binding heads of LacI as being distinguishable. Defining
$\alpha$ and $\beta$ as state variables that describe the
orientation of $O1$ and $Oid$ with respect to the binding heads,
respectively we can write the sum as
\begin{equation}
    \sum_i = \sum_{\mbox{heads}} \sum_{\alpha,\beta}.
\end{equation}
The sum over the heads results in a factor of two, since none of the
terms inside the sum actually depend on that choice. We next define
the overall looping energy $\Delta F_{\mbox{loop}}$ by
\begin{equation}
    e^{-\beta \Delta F_{\mathrm{loop}}} = {1 \over \sum_{\alpha,\beta} 1} \sum_{\alpha,\beta} e^{-\beta F_{\mathrm{loop},\alpha,\beta}}
            = {1 \over 4} \sum_{\alpha,\beta} e^{-\beta \Delta F_{\mathrm{loop},\alpha,\beta}}.
\end{equation}

Using the calculations and definitions from section
\ref{sec:LacISimple} we arrive at the looping probability
\begin{eqnarray}\label{eq:ploopStatMechNoStandard}
    p_{\mathrm{loop}} &=& \left[8 \frac{R}{\Omega} \frac{\delta \omega}{8 \pi^2} e^{-\beta(\Delta \varepsilon_1 +
                \Delta \varepsilon_{id} + \Delta F_{\mathrm{loop}} +2 \varepsilon_t - \varepsilon_{sol})}\right] \\
             &&\left[1+ 4 \frac{R}{\Omega} \frac{\delta \omega}{8 \pi^2} \left( e^{-\beta\Delta\varepsilon_1} +
                e^{-\beta\Delta\varepsilon_{id}}\right)+
                16 \frac{R(R-1)}{\Omega^2} \left(\frac{\delta \omega}{8 \pi^2}\right)^2
                e^{-\beta(\Delta\varepsilon_1+\Delta \varepsilon_{id})} + \right. \nonumber \\
             &&\left. 8 \frac{R}{\Omega} \frac{\delta \omega}{8 \pi^2} e^{-\beta(\Delta \varepsilon_1 +
                \Delta \varepsilon_{id} + \Delta F_{\mathrm{loop}} +2 \varepsilon_t - \varepsilon_{sol})}\right]^{-1}. \nonumber
\end{eqnarray}
Notice that the term that corresponds to looping has the energy
$\Delta F_{\mathrm{loop}}+2 \varepsilon_t - \varepsilon_{sol}$. In
principle this is the parameter associated with looping, but it also
includes information about the energetics of LacI when it is in
solution and when it has only one head bound to the DNA. However, we
can make the assumption that the energy associated with having half
a LacI in solution, $\varepsilon_t$ is half the energy of having a
full LacI in solution, $\varepsilon_{sol}$. This is equivalent to
saying that there is no change in the energetics of binding if the
other head is already bound, that there is no allosteric
cooperativity. If this is true then the parameter obtained from an
experiment where $p_{\mathrm{loop}}$ is measured will actually be
$\Delta F_{\mathrm{loop}}$.

Since we measure concentration of Lac repressor rather than absolute
number of repressor molecules we want to rewrite this formula as a
function of $[R]$ using the lattice definitions
\begin{equation}
    {R \over \Omega} = {R \over \Omega v} v = [R] v.
\end{equation}
The parameter $v$ corresponds to the volume of a lattice site, which
means that $\Omega v$ corresponds to the whole volume. We now make
the choice of a standard concentration
\begin{equation}
    {1 \over v} {8 \pi^2 \over \delta \omega} = 1~\mbox{M},
\end{equation}
which turns the looping probability from
eqn.~\ref{eq:ploopStatMechNoStandard} into
\begin{eqnarray}\label{eq:ploopStatMech}
    p_{\mathrm{loop}} &=& \left[8 {[R] \over 1~\mbox{M}} e^{-\beta(\Delta \varepsilon_1 +
                \Delta \varepsilon_{id} + \Delta F_{\mathrm{loop}})}\right] \\
             &&\left[1+ 4 {[R] \over 1~\mbox{M}} \left( e^{-\beta\Delta\varepsilon_1} +
                e^{-\beta\Delta\varepsilon_2}\right)+
                16 \left({[R] \over 1~\mbox{M}}\right)^2
                e^{-\beta(\Delta\varepsilon_1+\Delta \varepsilon_{id})} + \right. \nonumber \\
             &&\left. 8 {[R] \over 1~\mbox{M}} e^{-\beta(\Delta \varepsilon_1 +
                \Delta \varepsilon_{id} + \Delta F_{\mathrm{loop}})}\right]^{-1}. \nonumber
\end{eqnarray}

Finally, we make the connection to the thermodynamic formalism using
eqns.~\ref{eq:Kd2Epsilon} and by defining that
\begin{equation}\label{eq:J2Floop}
    J_{\mathrm{loop}} = \frac{1}{v} {8 \pi^2 \over \delta \omega} e^{-\beta \Delta F_{\mathrm{loop}}}.
\end{equation}
The point here is to use simple binding to define the parameters
$K_1$, $K_{id}$ and cyclization to assign the parameter
$J_{\mathrm{loop}}$ \citep{PBOC}. Here, we use a looping
$J_{\mathrm{loop}}$ factor rather than the regular factor $J$ factor
to emphasize the fact that the boundary conditions are different
from those present in cyclization, where $J$ is clearly defined
\citep{Shore1981}. In this way, we appeal to these other experiments
semantically and plug their definitions into the expression for the
looping probability derived above. This results in
\begin{equation}\label{eq:ploopThermo}
    p_{\mathrm{loop}} = \frac{{1 \over 2} \frac{[R]J_{\mathrm{loop}}}{K_1 K_{id}}}
                {1+\frac{[R]}{K_1} + \frac{[R]}{K_{id}} + \frac{[R]^2}{K_1 K_{id}} + {1 \over 2} \frac{[R]J_{\mathrm{loop}}}{K_1 K_{id}}},
\end{equation}
where $J_{\mathrm{loop}}$ is the average of the individual
$J_{\mathrm{loop}}$ factors over $\alpha$ and $\beta$ as defined in
eqn.~\ref{eq:Jtot}.

In the case where we distinguish between bottom and middle looped
states we can split $J_{\mathrm{loop}}$ into their corresponding
looping $J$ factors
\begin{equation}
    J_{\mathrm{loop}} = {1 \over 2} \left(J_{\mathrm{loop,B}} + J_{\mathrm{loop,M}} \right).
\end{equation}
In this case, for example, the probability of looping into the
bottom state can be written as
\begin{equation}\label{eq:ploopThermoIndiv}
    p_{\mathrm{loop}} = \frac{{1 \over 4} \frac{[R]J_{\mathrm{loop,B}}}{K_1 K_{id}}}
                {1+\frac{[R]}{K_1} + \frac{[R]}{K_{id}} + \frac{[R]^2}{K_1 K_2} + {1 \over 2} \frac{[R]J_{\mathrm{loop}}}{K_1 K_{id}}}.
\end{equation}

\subsection{Comparison of Theory and
Experiment}\label{sec:TheoryVsExperiment}

One of the important goals of this work is to demand a rich
interplay between theories of transcriptional regulation and
corresponding experiments.  To that end, the entirety of the data
presented in the paper is viewed through the prism of the
statistical mechanics model described above.

\begin{figure}
\begin{center}
\includegraphics[width=5in]{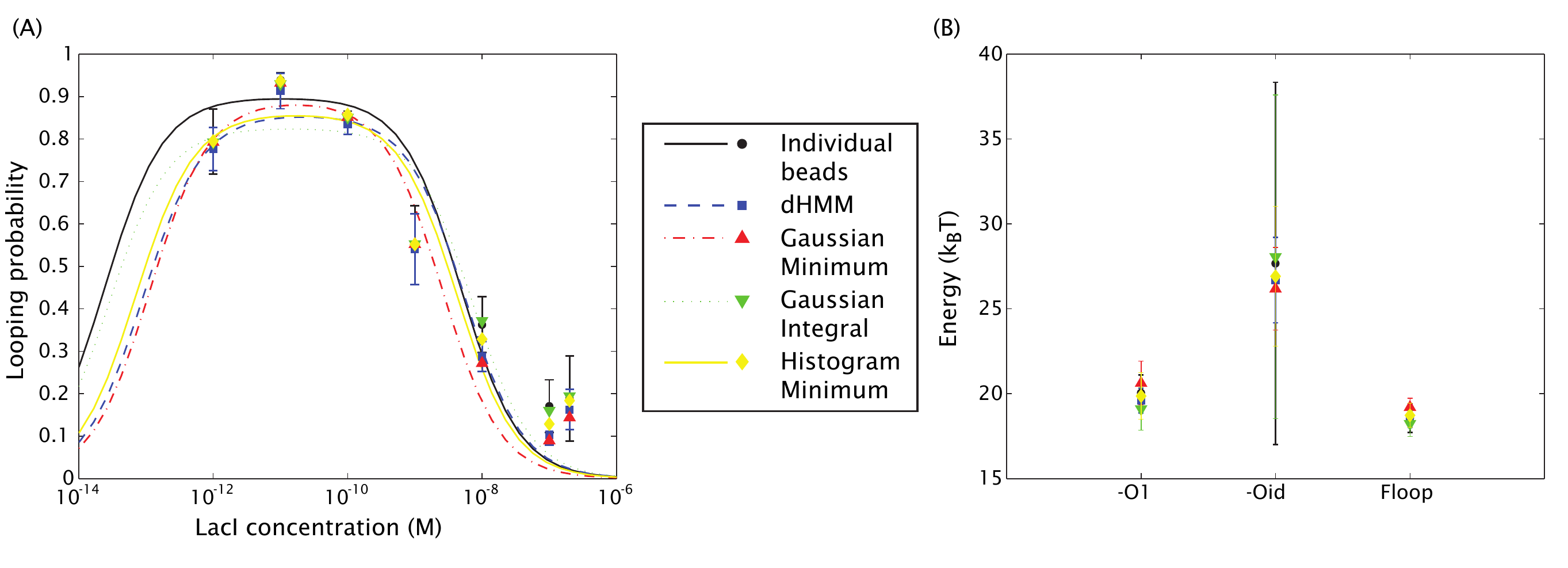}
\caption{\small{Alternative methods for fitting the looping probabilities.  (A) Different
schemes for determining the looping probability from the data result in
slightly different fits for the concentration dependent data.  (B) Results of the various
fits performed in (A). Notice how the model cannot constrain the binding energy of $Oid$ very accurately.}}
\label{LoopFitsHMMGaussian}
\end{center}
\end{figure}

One of the questions that we have examined is how the statistical
mechanics fit depends upon the choice of how we analyze the data to
determine the looping probability.    Examples of different schemes
for determining the looping probability and their allied fits are
shown in fig.~\ref{LoopFitsHMMGaussian}.
  In the main body of the paper, we presented looping probabilities
based upon Gaussian fits to the looping peaks.  However, we have
also explored the use of alternatives such as the Diffusive Hidden
Markov Model.

\begin{figure}
\begin{center}
\includegraphics[width=5in]{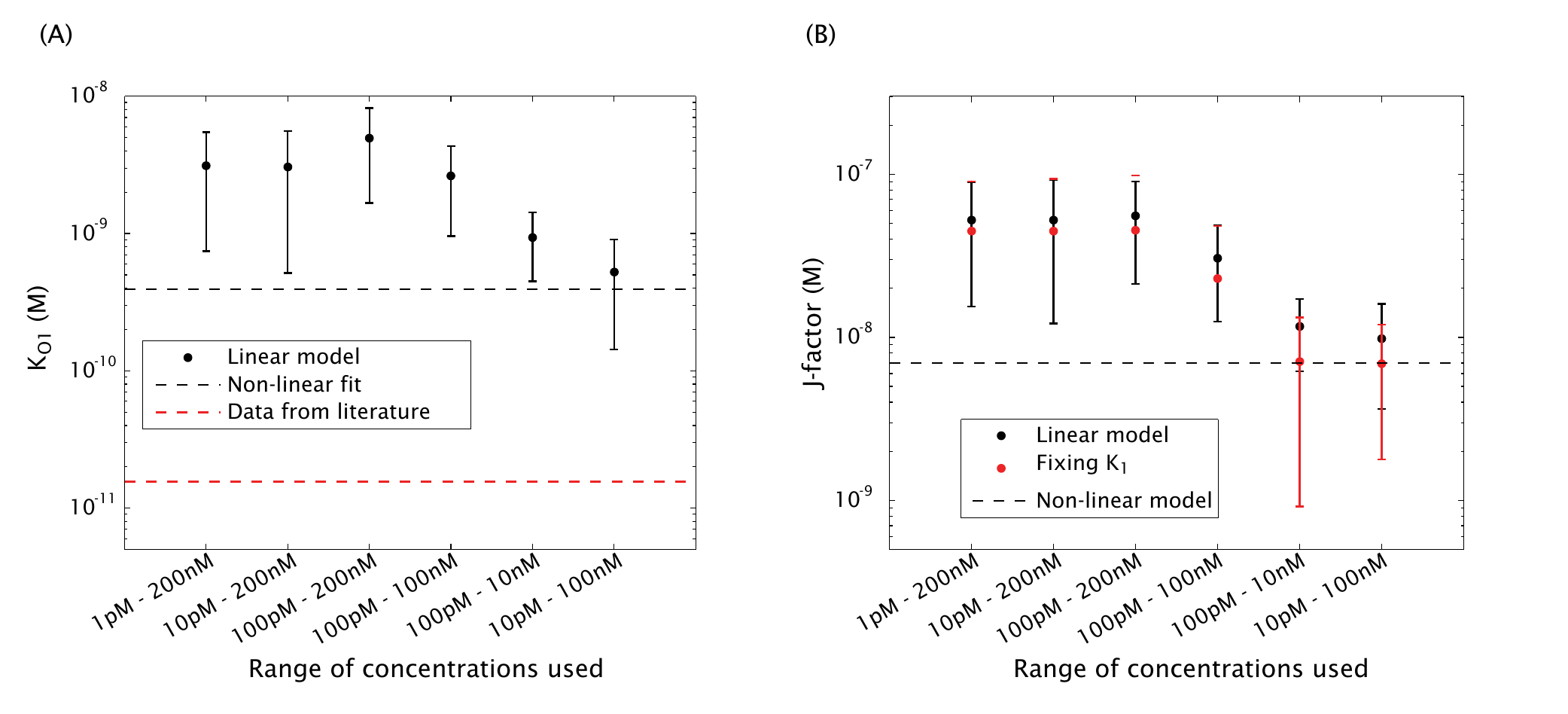}
\caption{\small{Sensitivity of fits to the  method of data analysis. (A) Different fits
to the value of $K_1$ using the linear model of eqn.~\ref{eq:pratioThermo1} and different ranges of data points from
fig.~\ref{fig:loopingpro}. The results corresponding to the non-linear model
of eqn.~\ref{eq:ploopThermo} are also shown. (B) Different fits to the value of $J_{\mathrm{loop}}$ using the linear and
non-linear models as shown in (A). ``Fixing $K_1$'' corresponds to fixing the $O1$ dissociation constant to the
literature value shown in table \ref{tab:FitTable}.}}
\label{FitResults}
\end{center}
\end{figure}

Another point of curiosity concerns the extent to which our fits for
the equilibrium constants and effective $J$-factor depends upon
which points from fig.~\ref{fig:loopingpro} are actually used to
make the fit.  Fig.~\ref{FitResults} shows the fit to both $K_1$ and
$J_{\mathrm{loop}}$ as a function of the particular model (nonlinear
or linear) and range of data points from fig.~\ref{fig:loopingpro}
that are used in the fit. The key observation is that the final two
data points (i.e. those at the largest concentrations of Lac
repressor) lead to a systematic shift in the values for both $K_1$
and $J_{\mathrm{loop}}$ when fitting using the linear model from
eqn.~\ref{eq:pratioThermo1}. Another interesting point revealed by
fig.~\ref{FitResults}(A) is that the full nonlinear model fit
results in a value for $K_1$ that is too large relative to the
literature value by roughly a factor of 10, corresponding to a
difference in binding energy of roughly 2~k$_B$T.

\begin{figure}
\begin{center}
\includegraphics[width=3in]{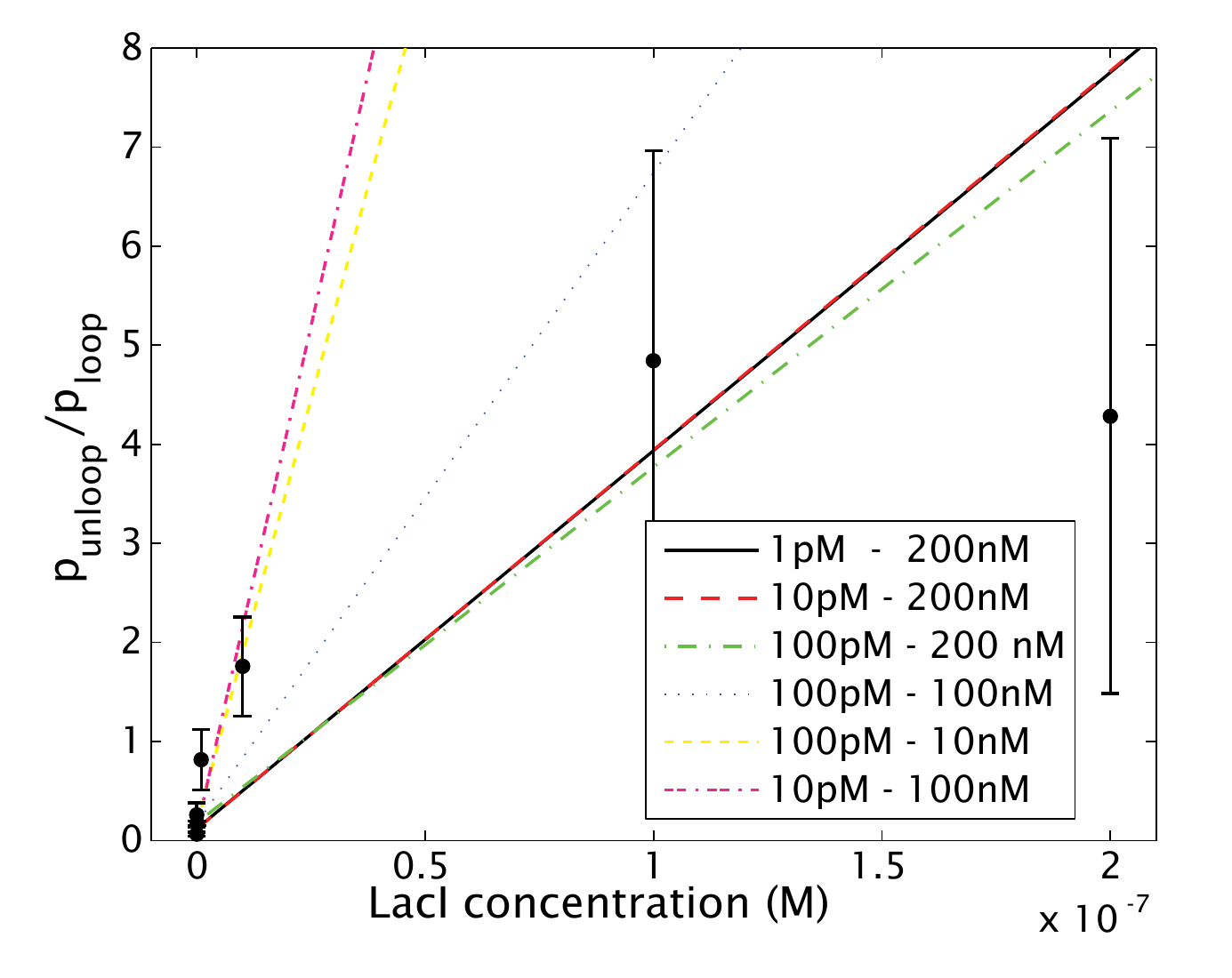}
\caption{\small{Sensitivity of linear fits to the range of data used. Different ranges of concentration from
fig.~\ref{fig:loopingpro} are fit using the linear model of eqn.~\ref{eq:pratioThermo1}.}}
\label{FitResults3}
\end{center}
\end{figure}

The dependence of our fits on the choice of data points included is
also revealed in fig.~\ref{FitResults3}.   In this case, we show the
result of using eqn.~\ref{eq:pratioThermo1} as the basis of the fit
and including different subsets of the data from
fig.~\ref{fig:loopingpro}.

\subsection{Individual Looped States}\label{sec:IndivLoops}

\begin{figure}
\begin{center}
    \includegraphics[height=2.0truein]{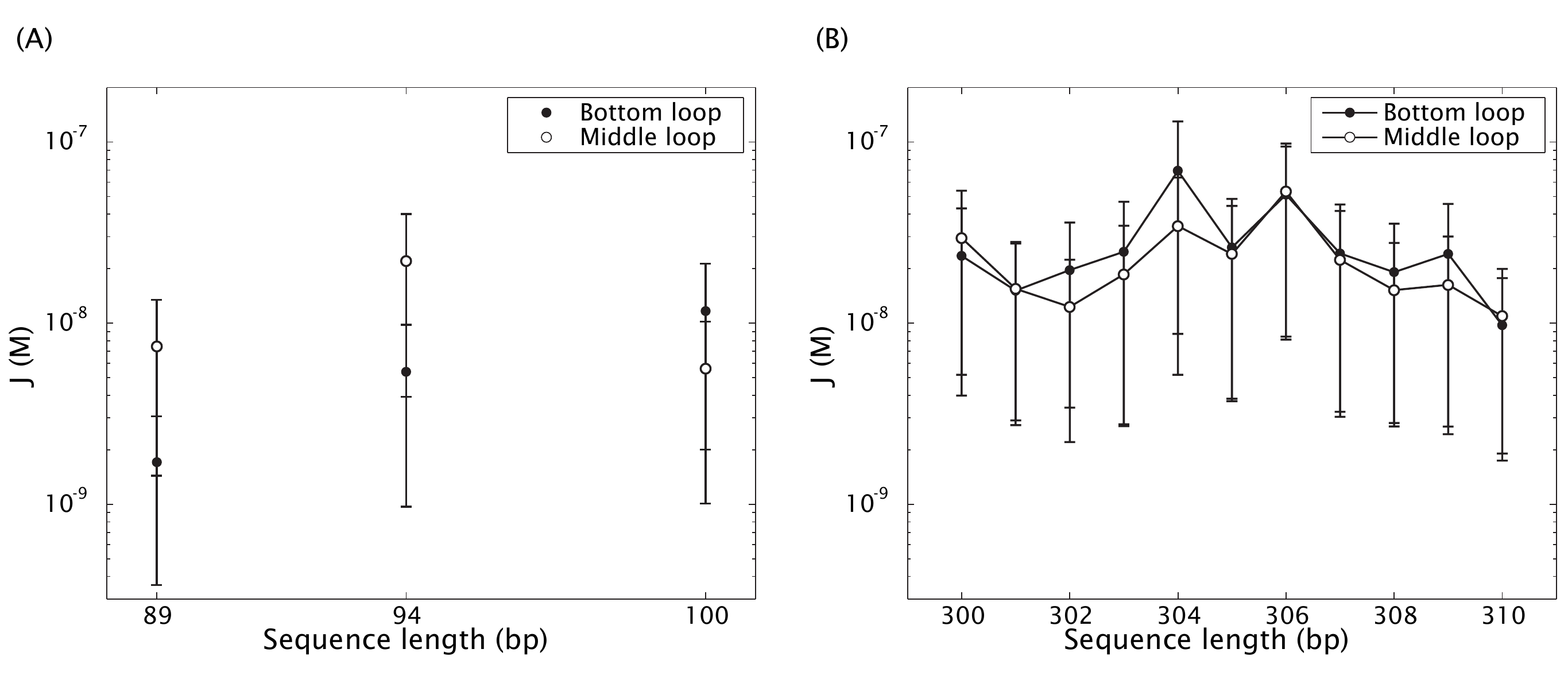}\\
    \caption{\small{Individual loops $J_{\mathrm{loop}}$ as a function of sequence length.
    (A) Results for short constructs, (B) results for long constructs.}}\label{fig:JLengthIndividual}
\end{center}
\end{figure}

\begin{figure}
\begin{center}
    \includegraphics[height=2.0truein]{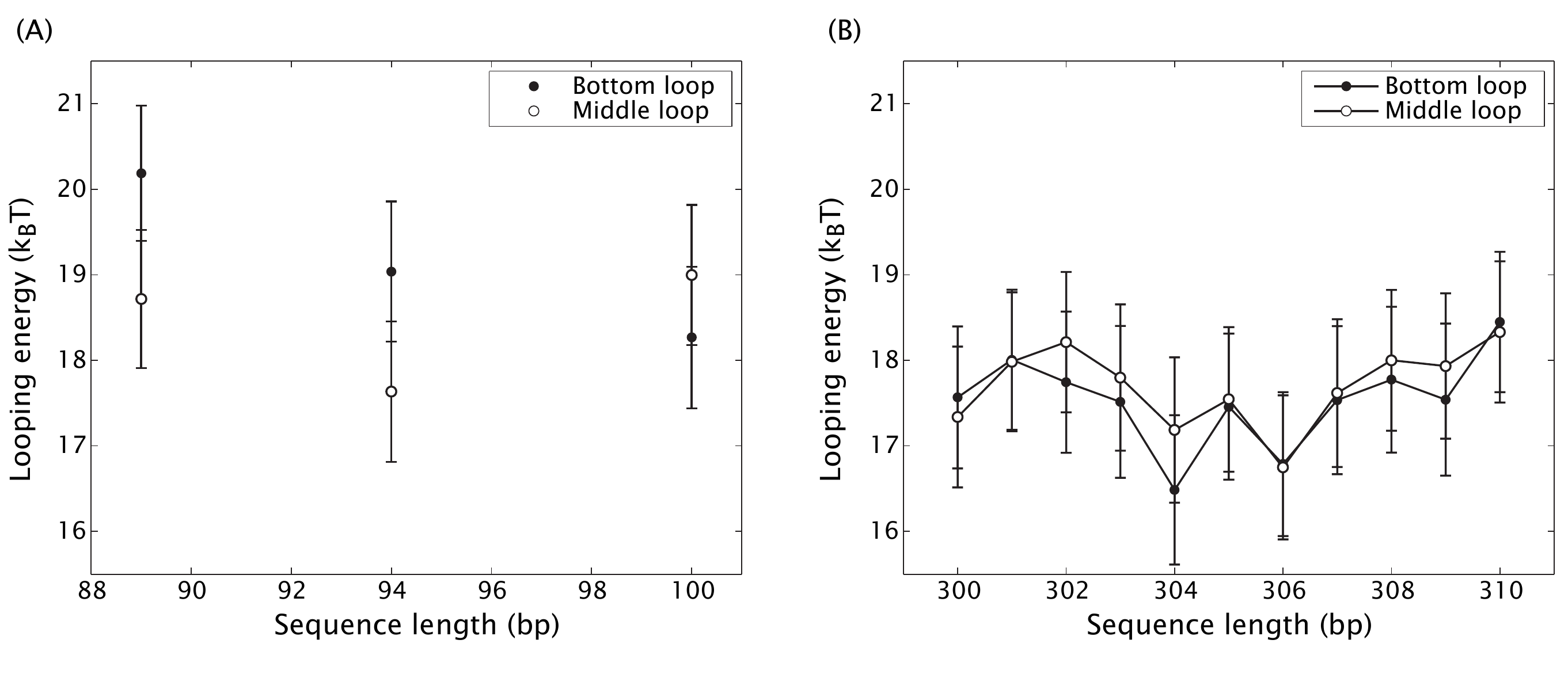}\\
    \caption{\small{Individual loops energies as a function of sequence length.
    (A) Results for short constructs, (B) results for long constructs.}
    }\label{fig:FloopLengthIndividual}
\end{center}
\end{figure}

In figs.~\ref{fig:pLoopConcIndividual} and
\ref{fig:pLoopLengthIndividual} we showed the looping probabilities
corresponding to each individual loop: the bottom and middle loops.
In order to analyze these results we can construct an individual
loop ratio analogous to the one defined in
eqn.~\ref{eq:pratioThermo1}. For the case of the bottom loop, for
example, this is
\begin{equation}
    p_{\mathrm{ratio,B}} = {p_{\mathrm{loop,B}} \over p_{\mathrm{unloop}}} =
        {4 K_1 \over J_{\rm loop,B}} + {4 [R] \over J_{\rm loop,B}}.
\end{equation}
Using an approach analogous to the one leading to
eqn.~\ref{eq:pratio2J} we obtain the looping $J$ factors associated
with each individual loop as shown in
fig.~\ref{fig:JLengthIndividual}. In
fig.~\ref{fig:FloopLengthIndividual} we show their corresponding
looping energies.

\clearpage

\subsection{Monte Carlo simulation\label{s:mcc}}

Our mathematical model built on our previous work
\citep{sega06a,nels06a,nels07a}, which showed that a
Gaussian-sampling simulation could accurately model the
experimentally observed relation between DNA tether length and TPM
bead motion by including an effective entropic stretching force from
bead--wall repulsion. This technique is essentially a Monte Carlo
evaluation of the equilibrium partition function of a chain. Instead
of a Metropolis implementation, we simply generated many discretized
chains using Gaussian distributions for each link's bending and
twisting angles, then discarded any such chains that violated the
global steric constraints. To compute looping $J$ factors, we
modified our previous code to monitor the separation and relative
orientation of the operator centers in the generated chains, and
found the fraction of all chains that met the conditions needed for
looping.   See~\citep{Towles2008} for more details.

To obtain the distributions of bead excursion shown in
fig.~\ref{fig:PNCombined}, we needed to make a correction before
comparing to the experimental data. Our video camera gathers light
for almost the entire $33\,$ms video frame time. This time scale is
an appreciable fraction of the bead's diffusion time in the trap
created by its tether, leading to a blurring of the bead image and
an apparent reduction of bead RMS excursion. We measured this effect
by looking at the apparent RMS excursion for a bead/tether system
with many different shutter times, then corrected our numerically
generated values for the position of the bead center to account for
blurring~ \citep{Towles2008}.

In addition, we reduced our simulation data in a way that parallels
what was done with the experimental data. The experiment takes data
in the form of a time series for the projected location of the bead
center (relative to its attachment), that is, $(x(t),y(t))$. We
found the length-squared of these position vectors, $R^2$, then
applied a Gaussian filter that essentially averaged over a 4-s
window. To simulate equilibrium averages in this context, we
harvested batches of $N_{\rm samp}$ independent simulated chains and
found the standard deviation of excursion within each batch. From
the resulting series of values for $R_{\rm RMS}=\sqrt{\langle
R^2\rangle_{N_{\rm samp}}}$, we made a histogram representing the
probability density function of $R_{\rm RMS}$. To choose an
appropriate value for $N_{\rm samp}$, we found a characteristic time
scale for bead diffusion from the time autocorrelation function of
$R_{\rm RMS}$, then divided the 4~s window into $N_{\rm samp}$ slots
corresponding to the larger of the frame time, 33~ms, or the bead
diffusion time~\citep{Towles2008}.

\bibliography{arXivLoopLengthSubmit}

\begin{thebibliography}{100}
\bibitem{calibration}
Motion of the bead is systematically characterized with various {DNA} lengths
  ranging from 200bp to 3 kbp. {Such} {DNA} is then interpolated using a second
  order polynomial function to served as a calibration curve. {From} this
  curve, for any given length {DNA} tether in that range, the amplitude of the
  motion of the {DNA} tethered bead can be evaluated.
\bibAnnoteFile{calibration}
\end{thebibliography}
\end{document}